\documentclass[a4paper,11pt]{article}
\pdfoutput=1
\usepackage{jheppub}
\usepackage{graphicx}
\usepackage{subfigure}
\usepackage{soul}

\newcommand{\beq}{\begin{equation}}
\newcommand{\eeq}{\end{equation}}
\newcommand{\bea}{\begin{eqnarray}}
\newcommand{\eea}{\end{eqnarray}}
\newcommand{\n}{\nonumber \\}
\newcommand{\sig}{\langle \sigma_{\rm a} v \rangle}
\newcommand{\rel}{\Omega_\chi h^2}

\def\m1{M_1}
\def\m2{M_2}
\def\m3{M_3}

\def\mz{m_Z}

\def\ma{m_{A}}

\def\gev{\,{\rm GeV}}

\def\to{\rightarrow}
\def\ccto{\chi_1^0 \chi_1^0 \rightarrow}
\def\xen{XENON-100~}

\title{Dark matter and Higgs bosons in the MSSM}

\author[a]{Tao Han,}
\author[a]{Zhen Liu}
\author[b]{and Aravind Natarajan}

\affiliation[a]{Pittsburgh Particle physics, Astrophysics, and Cosmology Center, \\
Department of Physics and Astronomy, University of Pittsburgh, \\
3941 O'Hara St., Pittsburgh, PA 15260, U.S.A.}
\affiliation[b]{McWilliams Center for Cosmology, Carnegie Mellon University, \\
Department of Physics, 5000 Forbes Ave, Pittsburgh PA 15213, U.S.A.}

\emailAdd{than@pitt.edu}
\emailAdd{zhl61@pitt.edu}
\emailAdd{anat@andrew.cmu.edu}

\abstract{
We investigate dark matter (DM) in the context of the minimal supersymmetric extension of the standard model (MSSM).
We scan through the MSSM parameter space and search for solutions that
(a) are consistent with the Higgs discovery and other collider searches;
(b) satisfy the flavor constraints from $B$ physics;
(c) give a DM candidate with the correct thermal relic density; and
(d) are allowed by the DM direct detection experiments.
For the surviving models with our parameter scan, we find the following features:
(1) The DM candidate is largely a Bino-like neutralino with non-zero but less than $20\%$ Wino and Higgsino fractions;
(2) The relic density requirement clearly pins down the solutions from the $Z$ and Higgs resonances ($Z,h,H,A$ funnels) and co-annihilations;
(3) Future direct search experiments will likely fully cover the $Z,h$ funnel regions, and $H,A$ funnel regions as well except for the ``blind spots'';
(4) Future indirect search experiments will be more sensitive to the CP-odd Higgs exchange due to its $s$-wave nature;
(5) The branching fraction for the SM-like Higgs decay to DM can be as high as $10\%$, while those from heavier Higgs decays to neutralinos and charginos can be as high as $20\%$. We show that collider searches provide valuable information complementary to what may be obtained from direct detections and astroparticle observations.
In particular, the $Z$- and $h$-funnels with a predicted low LSP mass should be accessible at future colliders.
Overall, the Higgs bosons may play an essential role as the portal to the dark sector.
}

\preprint{~~PITT-PACC 1303}

\begin{document}
\maketitle
\flushbottom

\section{Introduction}
\label{sec:intro}

Observations of the cosmic microwave background, gravitational lensing, clustering of galaxies, galactic rotation curves, etc.~have provided compelling evidence for the existence of Dark Matter (DM), which is likely to be of particle origin.
One of the best motivated candidates for DM is the Weakly Interacting Massive Particle (WIMP), a good example of which is the Lightest Supersymmetric Particle (LSP) (for reviews, see \cite{jungman_etal_1996,Bertone:2004pz,Drees:2012ji}).
If WIMPs exist in the Galaxy, they may be detected through direct search experiments \cite{dama,cogent,cresst,new_cdms,xenon10,xenon,lux,X1T}.
The DAMA experiment \cite{dama} has detected an annual modulation in the measured recoil spectrum at the $8.9 \sigma$ level,
consistent with the presence of WIMP DM in the Galaxy. More recently, the CoGeNT \cite{cogent}, CRESST \cite{cresst} and CDMS \cite{new_cdms} experiments have also obtained results that are consistent with low mass WIMP DM.
On the other hand, these results have been challenged by other experiments such as  XENON-10 \cite{xenon10},  XENON-100 \cite{xenon} and more recently TEXONO \cite{texono}, which have excluded the parameter space favored by the DAMA, CoGeNT, CRESST and CDMS experiments.
Complementary to the direct searches, indirect detection experiments include the Fermi gamma ray space telescope~\cite{fermi_dwarfs}, Alpha Magnetic Spectrometer~\cite{ams02}, Air Cherenkov Telescopes\cite{hess,magic,veritas}, and CMB experiments such as Planck \cite{planck}, and the Wilkinson Microwave Anisotropy Probe (WMAP)~\cite{wmap, wmap9}. The WMAP observations place a lower limit on the particle mass $m_\chi \gtrsim$ 10 GeV, assuming a velocity-independent annihilation cross section $\sig$ = 1 pb$\times$c \cite{galli_etal_2009, hutsi_etal_2011, galli_etal_2011, natarajan_2012, giesen_cmb, evoli_cmb}. The non-observation of gamma rays from DM annihilation in the nearby dwarf galaxies \cite{fermi_dwarfs, savvas} has been used to place constraints on the DM particle mass $m_\chi \gtrsim 40$ GeV, for neutralino annihilation to the $b \bar b$ channel with a velocity-independent cross section $\sig = $ 1 pb$\times$c, although these bounds would be relaxed with a more general analysis including the velocity-dependent contributions \cite{Cotta:2011pm,Fowlie:2011mb,Uncertainty,Tsai:2012cs}.

On the other hand, the LHC experiments have made a historic discovery of the long-sought-after Higgs boson
predicted by the Standard Model (SM). The experiments also show no evidence for Beyond-SM Higgs bosons, nor other new physics such as Supersymmetry (SUSY) etc.~with the current data, seemingly in favor of heavy colored sparticles~\cite{Feng:2000gh, Wells:2004di, gauginos, Giudice:2004tc, Giudice:2010wb}.
Several authors have studied the present LHC data and the implications for DM, as well as the possibility that future LHC data  will provide information to the DM puzzle
\cite{Farina:2011bh,AlbornozVasquez:2011aa,Kadastik:2011aa,Bottino:2011xv,JEllis,baer,Cao:2012fz,Cao:2012im,Choudhury:2012tc,Baer:2012uy,belanger_etal,arbey_etal_1,Cao:2012yn,baer1,Allahverdi:2012wb,Mohanty:2012ri,Baer:2012se,Hisano:2012wm,Hall:2012zp,Altmannshofer:2012ks,CahillRowley:2012kx,arbey_etal_2,strege_etal_2012,Kowalska:2013hha}.
Although the SUSY parameter space has been significantly reduced due to the absence of a SUSY signal at the LHC and due to the constraining properties of the SM-like Higgs boson, a dark matter candidate can still be readily accommodated in SUSY theories.

With the ever increasing experimental sensitivity of DM detection experiments, we are motivated to explore to what extent DM properties have been constrained by the results
from particle accelerator experiments.
Our goal is to systematically examine the complementarity between DM direct detection experiments, indirect detection searches, and collider experiments, and in particular explore the potential pivotal role played by the Higgs bosons.
We perform a comprehensive study in the framework of the minimal supersymmetric extension of the standard model (MSSM).
We impose the following  constraints on our model considerations:
\begin{itemize}
\item[(1)]{{\bf Relic abundance:} the neutralino LSP constitutes all the cold DM, consistent with the cosmological observations~\cite{wmap, wmap9}.}
\item[(2)] {{\bf Collider constraints:}
the MSSM parameter space satisfies all collider constraints from the Higgs boson searches and has a SM-like Higgs boson near 126 GeV.}
\item[(3)] {{\bf Flavor constraints:} the parameter space satisfies the flavor constraints from
$b\rightarrow s\gamma$~\cite{Amhis:2012bh}},
$B_{\rm s}\rightarrow\mu^+\mu^-$~\cite{lhcb}.
\end{itemize}
We further check the consistency of the annihilation rate at zero velocity $\langle \sigma_{\rm a} v \rangle (v \rightarrow 0)$ with CMB observations, and the absence of gamma rays from nearby dwarf galaxies \cite{fermi_dwarfs, savvas}.
It is known that the spin-independent WIMP-nucleon elastic scattering cross section obtained by the XENON-100 experiment \cite{xenon} puts a very strong bound on the MSSM parameter space. We find that the surviving region has characteristic features, notably a Bino-like LSP.
What is most interesting to us is that all these scenarios would lead to definitive predictions for the LHC experiments, that can be verified by the next generation of direct/indirect search experiments such as LUX \cite{lux} and XENON-1T \cite{X1T}.

The rest of the paper is organized as follows.
In Section \ref{freeze_out}, we compute the WIMP relic density in a model-independent manner. We emphasize the importance of including the effect of velocity-dependent annihilation $\sig = a + b v^2 + \mathcal{O}(v^4)$, which is crucial when the $b$ term from the $p$-wave is not negligible. This will have direct consequences in the interpretation of indirect search results.
In Section \ref{sec:para}, we discuss our technique for scanning the MSSM parameter space. In Section 4, we present our results, and discuss the experimental constraints from the Higgs and flavor searches. We also discuss the constraints on the parameter space imposed by the XENON-100 search for spin-independent scattering, as well as the Super-K and IceCube/DeepCore limits on spin-dependent scattering. We show that future experiments such as LUX and XENON-1T  will likely probe the natural supersymmetric parametric space consistent with the LSP constituting all the DM.
We present extensive discussions of our results in Section \ref{Discuss} and finally draw our conclusions in Section \ref{Conclude}.
Details of the relic density calculation are presented in the Appendix.
%


\section{Dark matter relic density}
\label{freeze_out}

\begin{figure}[t]
\hspace{0.25in} (a) \hspace{2.7in}  (b)
\begin{center}
\scalebox{0.39}{\includegraphics{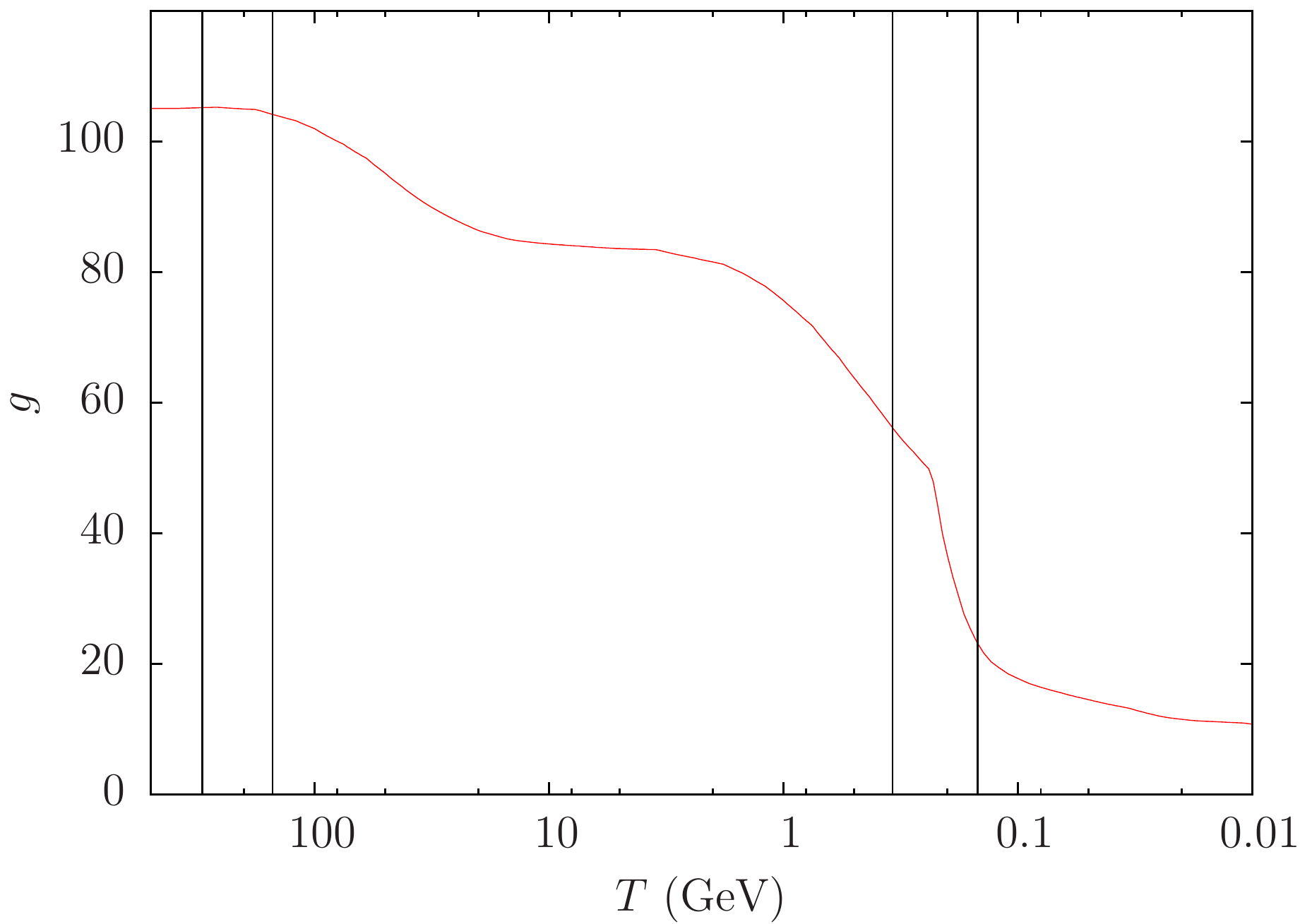}}
\put(-174,25){\rotatebox{-270} { {\footnotesize Electroweak transition}}}%
\put(-62,25){\rotatebox{-270} { {\footnotesize Quark-Hadron transition}}}%
\scalebox{0.39}{\includegraphics{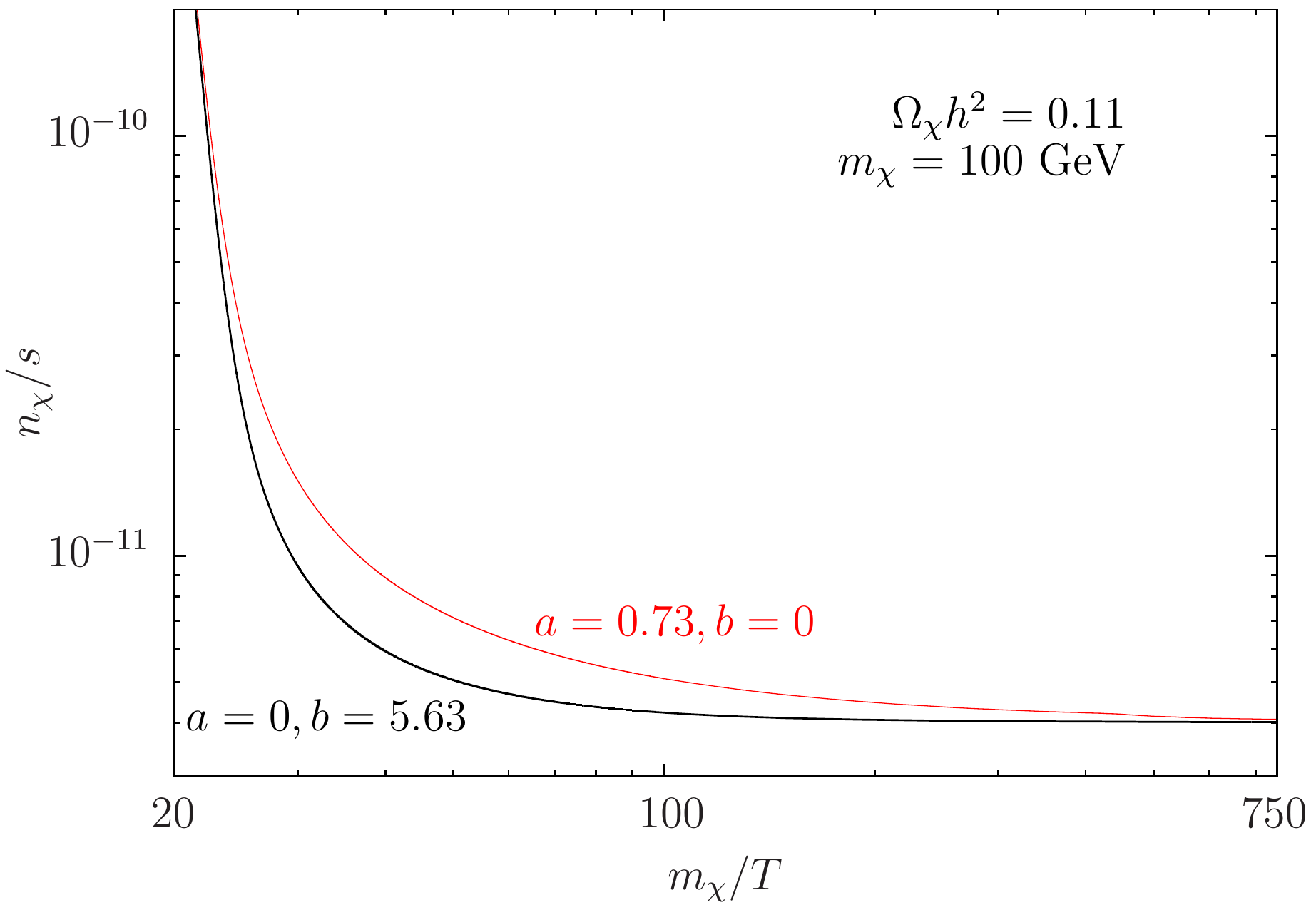}}
\end{center}
\hspace{0.25in} (c) \hspace{2.7in}  (d)
\begin{center}
\scalebox{0.39}{\includegraphics{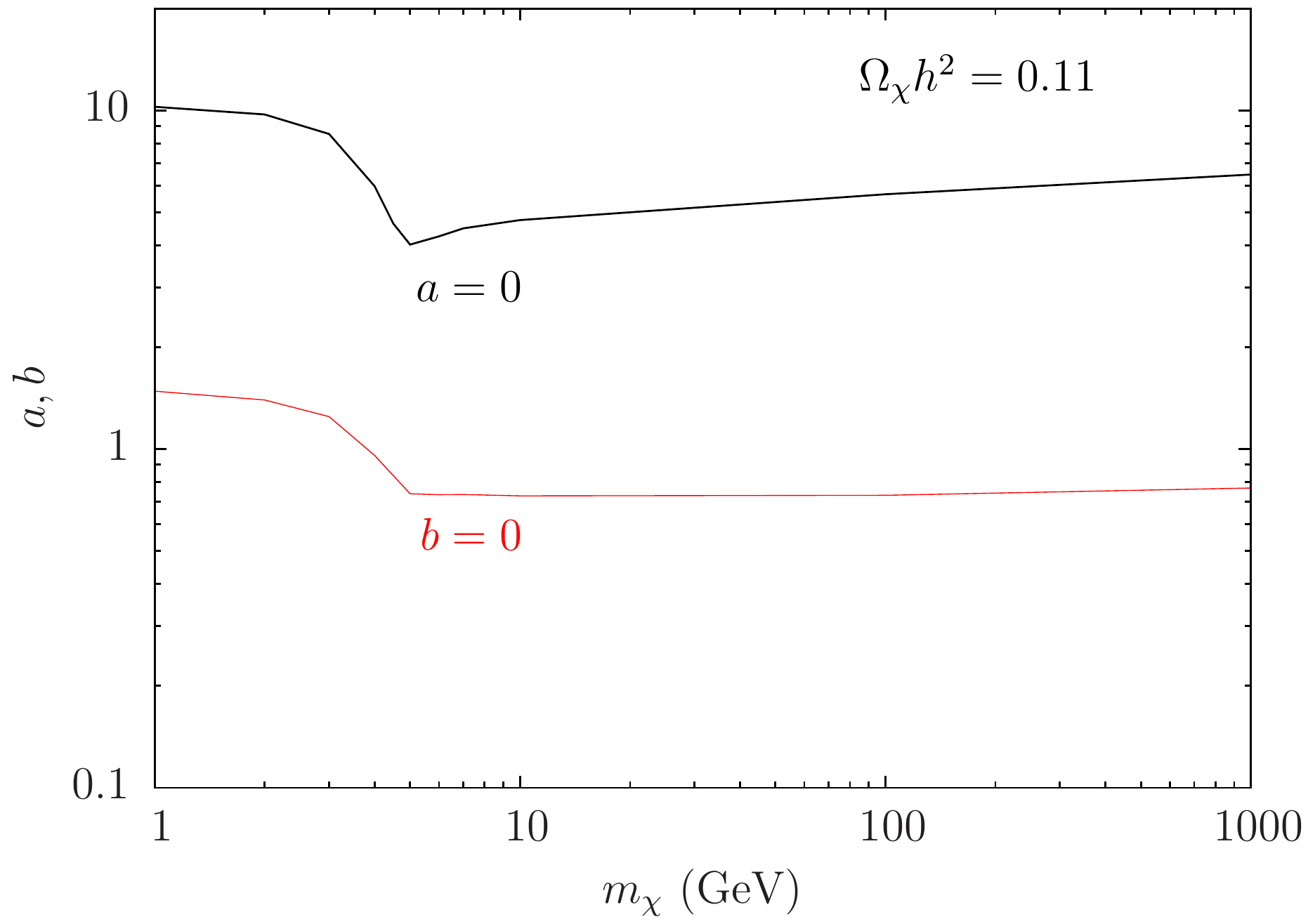}}
\scalebox{0.38}{\includegraphics{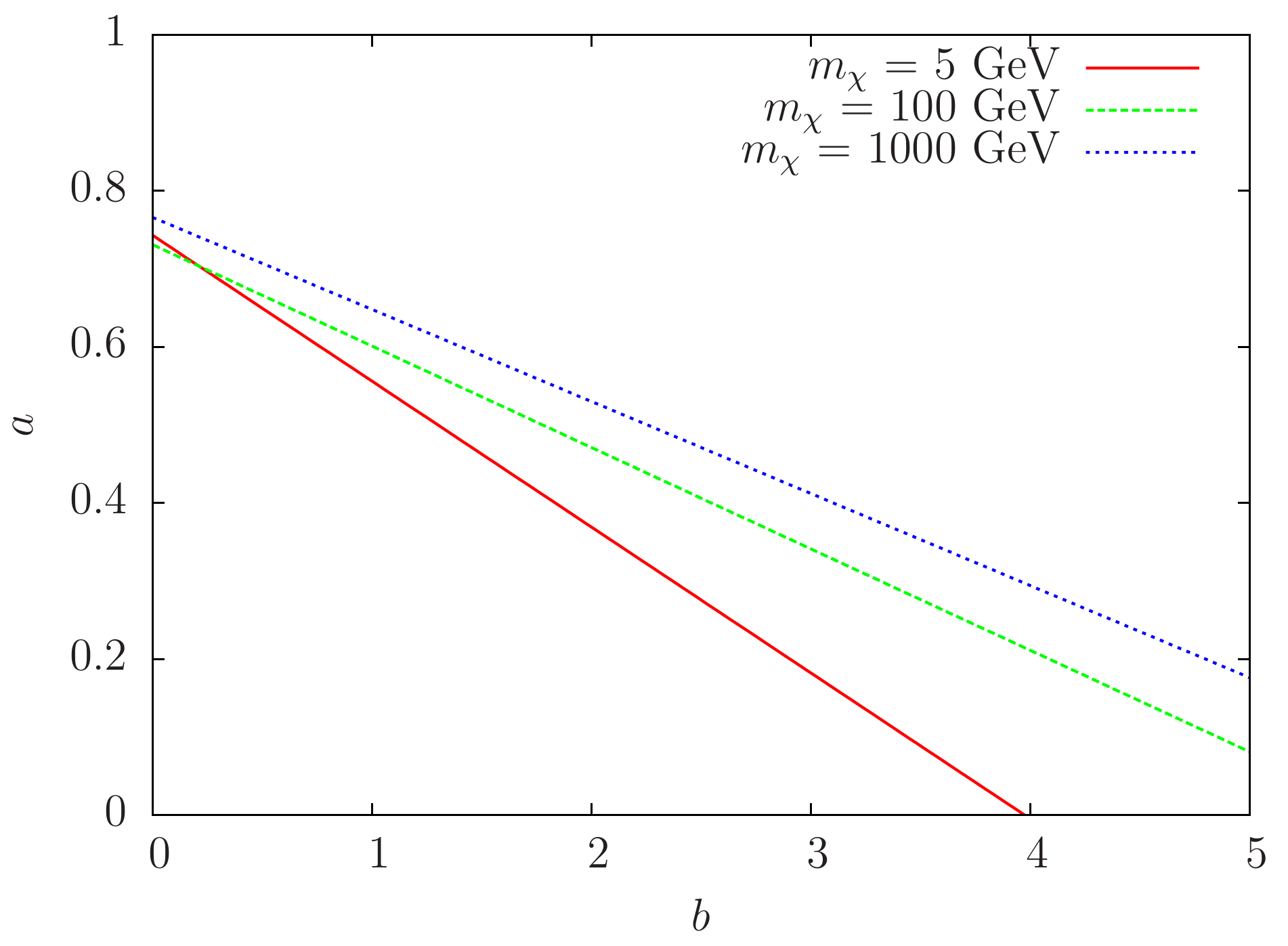}}
\end{center}
\caption[]{
(a) The number of relativistic degrees of freedom as a function of temperature;
(b) WIMP number density evolution with temperature for illustrative values of $a,b$ with $m_\chi = 100$ GeV;
(c) The coefficients $a$ and $b$ values for different WIMP masses that saturate the DM relic density;
(d) The coefficients $a$ versus $b$ for $m_\chi =$5, 100, and 1000 GeV.
The dark matter density fraction at the present epoch is set to $\Omega_\chi h^2$ = 0.11.
\label{fig1}}
\end{figure}

Within the Standard Cosmology, we evaluate the thermal history of the dark matter \cite{jungman_etal_1996,Bertone:2004pz}. We assume that the WIMP, generically denoted by $\chi$, constitutes all of the thermal DM. Let us define the DM relic density $\Omega_\chi$ as the ratio of the DM mass density at the present epoch to the critical mass density ($\rho_{\rm crit}$). The recent data from Planck implies a relic density
\beq
\rel = \frac{m_\chi n_{\chi}}{  \rho_{\rm crit} / h^2 }  =   0.1187 \pm 0.0017, 
\label{eq:relic}
\eeq
when combined with other CMB experiments and BAO observations. The number density of WIMPs at time $t$, $n_\chi (t)$, can be obtained by solving the Boltzmann equation (see Appendix for details).
The annihilation cross section ($\sigma_{\rm a}$) characterizes the WIMP dynamics for a given theory. Since WIMPs are non-relativistic at freeze-out, the velocity averaged cross section $\sig$ may be expanded in $v$, customarily written as
\beq
{ \sigma_{\rm a} v \over 1~{\rm pb} \times c } =   a + b v^2 + \mathcal{O}(v^4) , \quad
{ \sig \over 1~{\rm pb} \times c }
=   a + {6b \over x} \ \ {\rm with}\ \ x= {m_{\chi} \over T},
\label{eq:master}
\eeq
where the traditional units are 1 pb$\times c  = 3 \times 10^{-26}\ {\rm cm^3/s}$. Simple threshold arguments indicate that $s$-wave annihilation contributes dominantly to $a$, while $p$-wave annihilation contributes only to $b$. Requiring that the thermal relic density satisfy the measured value gives us the cross section. This leads to the result:
\beq
\rel \approx 0.11\  \Rightarrow \ \sig \approx  2.18 \times 10^{-26}\ {\rm cm^3/s}.
\eeq
Figure~\ref{fig1}(a) shows the evolution of the number of relativistic degrees of freedom, following \cite{laine_schroder_2006}, which change rapidly at $T \sim 150$ MeV due to the quark-hadron transition.
We show the WIMP number density approaching the present day value in figure~\ref{fig1}(b) for $m_{\chi}=100$ GeV for the extreme cases $b=0$ (pure $s$-wave annihilation), and $a=0$ (pure $p$-wave annihilation). The range of  $a$ and $b$ values  that saturate the DM relic density, versus the WIMP mass is shown in  figure~\ref{fig1}(c).
For a WIMP mass above around $4-5$ GeV, $a$ remains almost constant for arbitrary WIMP masses when $b$ is of negligible value, reflecting the ``WIMP miracle" that leads to the correct relic density. Below $4-5$ GeV, somewhat larger values of $a$ and $b$ would be needed to yield the correct relic density, due to the reduction of $g$ at the quark-hadron transition. We will not explore the very low mass region any further in this work. The interplay between $a$ and $b$ follows a linear relation empirically, and is shown in figure~\ref{fig1}(d) for various WIMP mass values,
consistent with a present day dark matter relic density.

\section{The MSSM parameters  relevant to DM studies}
\label{sec:para}

In SUSY theories with conserved R-parity, the lightest supersymmetric particle (LSP) is a viable WIMP DM candidate. For both theoretical and observational considerations \cite{jungman_etal_1996,Bertone:2004pz, Neutralino1, Neutralino3, Neutralino2}, it is believed that the best candidate is the lightest Majorana mass eigenstate which is an admixture of the Bino ($\tilde{B}$), Wino ($\tilde{W_3}$), and Higgsinos ($ \tilde{H}_{d,u}$), with the corresponding soft SUSY breaking mass parameters $M_{1},\ M_{2}$, and the Higgs mixing $\mu$, respectively. The neutralino mass matrix in the Bino-Wino-Higgsino basis is given by
\[
M_{\rm neut} = \left [
  \begin{array}{cccc}
  M_1 & 0 & -m_{\rm z} \cos\beta \sin\theta_{\rm w} & m_{\rm z} \sin\beta \sin\theta_{\rm w}\\
  0 & M_2 & m_{\rm z} \cos\beta \cos \theta_{\rm w} & -m_{\rm z} \sin\beta \cos \theta_{\rm w} \\
  -m_{\rm z} \cos\beta \sin\theta_{\rm w} &  m_{\rm z} \cos\beta \cos \theta_{\rm w} & 0 & -\mu \\
  m_{\rm z} \sin\beta \sin\theta_{\rm w} & -m_{\rm z} \sin\beta \cos \theta_{\rm w} & -\mu & 0 \\
  \end{array}
  \right ],
\]
where $m_{\rm z}$ is the $Z$ boson mass, $\theta_{\rm w}$ the Weinberg angle, and $\tan\beta=v_{u}/v_{d}$ is the ratio of the vacuum expectation values for the two Higgs doublets. The lightest neutralino is a linear combination of the superpartners
\beq
\chi^0_{1} = N_{11} \tilde{B} + N_{12} \tilde{W_3} + N_{13} \tilde{H_d} + N_{14} \tilde{H_u},
\eeq
where $N_{\rm ij}$ are the elements of the matrix $N$ that diagonalize $M_{\rm neut}$:
\beq
 N^* M_{\rm neut} N^{-1}  = {\rm diag} \{ m_{\chi^0_1}, m_{\chi_2^0}, m_{\chi_3^0}, m_{\chi_4^0} \}.
 \eeq
The eigenvalues of $M_{\rm neut}$ are the masses of the four neutralinos.
An interesting limit is $m_{\rm z} \ll |M_{1}\pm \mu|$ and $|M_{2}\pm \mu|$, in which case, the mass eigenstates
(neutralinos $\chi^{0}_{i}$) are nearly pure gauge eigenstates (gauginos and Higgsinos).
This also implies that large mixing of gaugino and Higgsino components for the mass eigenstates only takes place when $M_1$ and/or $M_2$ are nearly degenerate with $\mu$.
We will focus only on the lightest neutralino (henceforth denoted by $\chi_1^{0}$) with a mass $m_{\chi_1^0}$. In particular, we assume that it constitutes the majority of the DM.

Intimately related to the neutralinos is the Higgs sector. The tree level Higgs masses in the MSSM can be expressed in terms of $\tan\beta$ and the CP-odd mass $M_{A}$. Radiative corrections enhance the Higgs mass significantly via the top quark Yukawa coupling, the third generation squark mass parameters $M_{Q3}, \ M_{U3}$, and the left-right squark mixing $A_{\rm t}$. Flavor physics observations from the $b$-quark sector often serve as stringent constraints and we therefore include the sbottom sector parameters $M_{D3}$ and the squark mixing $A_{\rm b}$. The last potentially relevant sector is the stau, which could be light and contribute to the $t$-channel exchange, co-annihilations to control the relic density.
We therefore generously vary the MSSM parameters in the ranges
\bea
\nonumber
5 \gev < |M_1| < 2000 \gev,&\ \  100 \gev <& |M_2,~\mu| < 2000\gev, \\ \nonumber
\label{eq:scan}
3 < \tan\beta < 55,~~~~~&\ \  80 \gev <& M_{\rm A} < 1000 \gev, \\
-4000 \gev < A_{\rm t} < 4000 \gev,& 100 \gev <& M_{\rm Q3},~M_{\rm U3} < 3000 \gev, \\ \nonumber
-4000 \gev < A_{\rm b} < 4000 \gev,& 100 \gev <& ~~~M_{\rm D3}~~~~~ < 3000 \gev, \\ \nonumber
-4000 \gev < A_{\rm \tau} < 4000 \gev,& 100 \gev <& M_{\rm L3},~M_{\rm E3} < 3000 \gev.
\nonumber
\eea
The lowest values of $M_1$, $M_2$ and $\mu$ control the LSP mass for the WIMP DM. The lower values of 100 GeV for $M_2,\ \mu$ are dictated by the LEP-2 bound from the largely model-independent chargino searches.
The lower limit of $\tan\beta$ is close to the LEP-2 Higgs search exclusion.
The lower limit of $M_{A}$ is chosen to cover the non-decoupling Higgs sector as well as above the LEP-2 bound on the charged Higgs.
The upper limit of $M_1$, $M_2$, $\mu$ and the soft SUSY breaking masses in the stop and stau sectors are set with  consideration of naturalness \cite{Barbieri:1987fn,Kitano:2006gv,barger_langacker_etal, Baer:2012uy,CahillRowley:2012rv,Feng:2013pwa}.
The other soft supersymmetry breaking parameters are less relevant for our DM considerations and we therefore set the other trilinear mass parameters to be zero, and the other soft SUSY breaking masses at 3 TeV.

While the natural value of $\mu$ is supposed to be close to the electroweak scale, we vary $\mu$ up to 2 TeV to capture some interesting features such as the scenario of ``well-tempered neutralino'' \cite{ArkaniHamed:2006mb}. Letting $\mu \approx$ 2 TeV would already allow for a severe fine tuning at the level of about
$0.04\%$ \cite{Barbieri:1987fn}. Although not our focus, we have included arbitrary signs for the $M_1$, $M_2$, $\mu$ parameters. This allows us to see the possible solutions with very specifically chosen parameter relations such as the
 ``blind spots'' scenarios \cite{Perelstein:2011tg,Grothaus:2012js,Cheung:2012qy}.

We choose a flat prior for the scanning with a total number of scanned points around $10$ million. Several different layers of scanning are performed to account for different experimental constraints and theoretical considerations, as seen by the corresponding color codes in our plots.


%
\section{Current constraints and the scanning results}

The hints of DM detection from the DAMA, CoGeNT, and CRESST experiments have drawn significant interest in considering valid theoretical interpretations.
The sensitivity of the DM direct searches have been steadily improving at an impressive pace, notably with the XENON collaboration \cite{xenon10,xenon}. The indirect searches from WMAP, Fermi-LAT, and IceCube have also played crucial roles in exploring the nature of the DM particle.

Although the null results of searching for Supersymmetry at colliders have significantly tightened the viable SUSY parameter region, the bounds on WIMP DM properties are only limited within specific models, most notably in mSUGRA or CMSSM \cite{baer,JEllis}. The direct exploration of the electroweak gaugino sector at the LHC would be very challenging given the hostile background environment and the current search results depend on several assumptions\cite{Chatrchyan:2012pka,SanjayHanSu}.
On the other hand, if we demand the correct WIMP LSP relic abundance from the current observations as in eq.~(\ref{eq:relic}), the SUSY parameter space of eq.~(\ref{eq:scan}) will be notably constrained in the Higgs and gaugino sectors. We assume a $10\%$ theoretical uncertainty in the calculation of the DM relic density \cite{Uncertainty,Akcay:2012db}. Applying the Planck result for $\Omega_\chi h^2$
in eq.~(\ref{eq:relic}) combined with $10\%$ theoretical uncertainty, we demand that the relic density in our model points be within the following $2\sigma$ window
\beq
\centering
0.0947~<~\Omega_{\chi_1^0}h^2~<~0.1427~.   
\label{eq:relic_n}
\eeq
We use the publicly available {\scriptsize FEYNHIGGS} code \cite{FeynHiggs} as the spectrum calculator. The Higgs constraints are imposed using the {\scriptsize HIGGSBOUNDS} package \cite{HiggsBounds} with our additional modifications. We modify the codes to include the most recent LHC constraints \cite{CMS-PAS-HIG-12-044,CMS-PAS-HIG-12-043,CMS-PAS-HIG-12-042,CMS-PAS-HIG-12-041,CMS-PAS-HIG-12-045,ATLAS-CONF-2012-170,ATLAS-CONF-2012-169,ATLAS-CONF-2012-168,ATLAS-CONF-2012-163,ATLAS-CONF-2012-162,ATLAS-CONF-2012-161,ATLAS-CONF-2012-160,ATLAS-CONF-2012-158}.
%
The standard SLHA \cite{SLHA} output recorded is then supplied to the {\scriptsize MICROMEGAS} code \cite{MicrOmegas} which computes the DM relic density, direct/indirect search cross sections and flavor calculations. This is done to avoid any possible inconsistency due to the subtle differences in the spectrum calculator, particularly the lack of accuracy in the default approximate diagonalization routine for the neutralino mass matrix.
%


\subsection{Constraints from the Higgs searches and the flavor sector}

The discovery of a SM-like Higgs boson $h$ as well as the upper limits on difference channels for the other Higgs bosons
$A,H,H^{\pm}$ shed much light on the electroweak sector, and can thus guide us for DM studies.
When scanning over the SUSY parameter space as in eq.~(\ref{eq:scan}), and requiring the correct WIMP LSP relic abundance
to be within the $2\sigma$ window in eq.~(\ref{eq:relic_n}), we further require the theory to have a SM-like Higgs boson, and to accommodate all the current constraints from the Higgs searches:
\bea
\nonumber
&&123\gev < m_h < 128\gev,\quad \sigma_{\gamma\gamma} > 0.8\ \sigma_{\gamma\gamma}(SM), \\
\label{eq:requirement}
&&{\rm plus\ Higgs\ search\ bounds\ from\ LEP,\ Tevatron,\ LHC,}\\ \nonumber
&&{\rm plus\ LEP\ bounds\footnotemark[1]\ on\ the\ slepton\ mass\ (\ge~80~\gev)}\\ \nonumber
&&{\rm and\ the\ squark\ and\ the\ chargino\ mass\ (\ge~100~\gev).}\nonumber
\eea
The Higgs diphoton rate being SM-like is one of our assumptions. We do not demand it to reach a large excess as indicated by the early LHC results, nor do we accept the deficit as suggested by the latest CMS results \cite{ATLAS-CONF-2013-012,CMS-PAS-HIG-13-001}. It is a statement of having a SM-like Higgs boson.
Due to the correlation of the Higgs couplings, the requirement of the $\sigma_{\gamma\gamma}$ cross section effectively sets the SM-like values for $\sigma_{WW},\ \sigma_{ZZ}$ as well.
\footnotetext[1]{The particle mass constraints applied here may still be evaded for certain limiting cases, if the lower lying particles have a mass splitting less than the order of GeV, for instance.}

The absence of tree-level flavor changing neutral currents (FCNC) in the SM puts strong constraints on new physics. We consider two processes that have been observed to be consistent with the SM prediction and thus provide constraints on the MSSM parameter space.
The first process is $b \rightarrow s \gamma$ \cite{bsg1}, for which the branching fraction is sensitive to the charged Higgs boson and supersymmetric particles (e.g.~chargino/stop) in the loop. The world average of the branching fraction of this channel~\cite{Amhis:2012bh}
is
$\left( 3.43 \pm 0.21\pm 0.07 \right ) \times 10^{-4}$, in good agreement with the standard model prediction \cite{Misiak:2006zs,Becher:2006pu,Benzke:2010js} $\left ( 3.15\pm0.23 \right ) \times 10^{-4}$.

The second process is $B_{\rm s} \rightarrow \mu^+\mu^-$, which receives a large contribution in the MSSM  proportional to  $\left ( \tan^6 \beta / m^4_{\rm A} \right )$ \cite{Babu:1999hn}. The LHCb collaboration has recently announced the first evidence \cite{lhcb} of this very rare decay
and the branching ratio for this process was found to be $\left( 3.2 ^{+1.4~+0.5} _{-1.2~-0.3} \right ) \times 10^{-9}$ in good agreement with the standard model prediction of  $ \left( 3.23 \pm 0.27 \right ) \times 10^{-9}$ \cite{Buras:2012ru}. We adopt world average of the branching fraction of this channel~\cite{Amhis:2012bh} $ \left( 3.2 \pm 1.0 \right ) \times 10^{-9}$ to put constraints on BR($B_{\rm s} \rightarrow \mu^+\mu^-$).
\begin{figure}[t]
\begin{center}
\subfigure[]{
      \includegraphics[width=209pt]{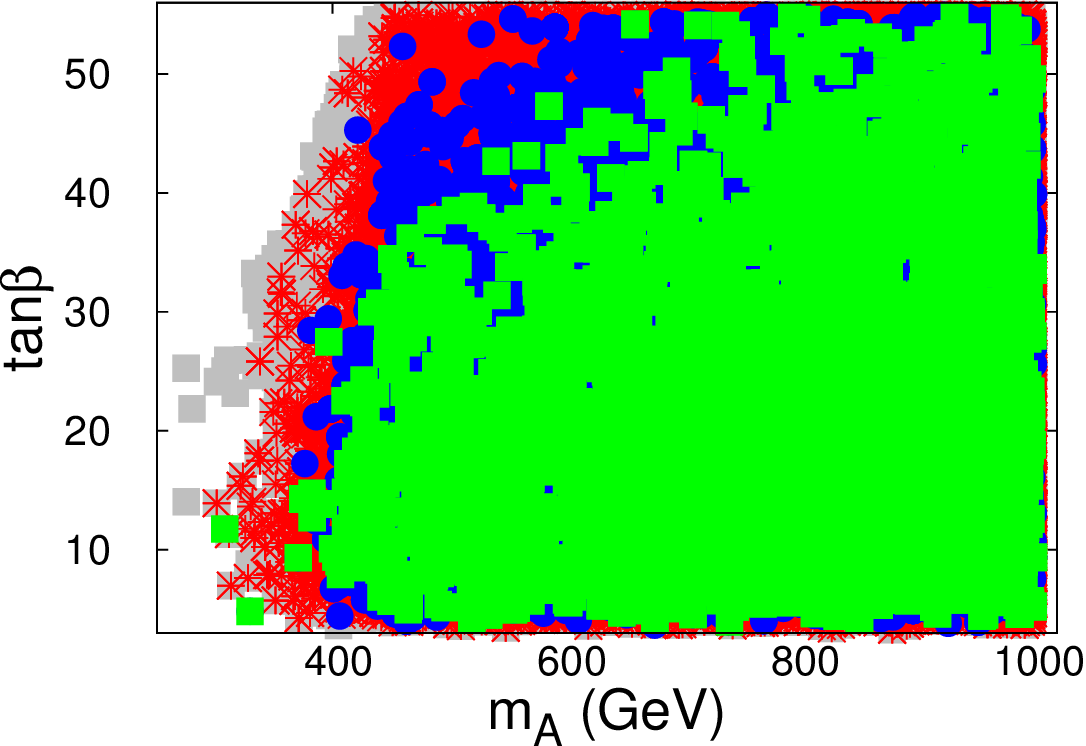}}
\subfigure[]{
      \includegraphics[width=209pt]{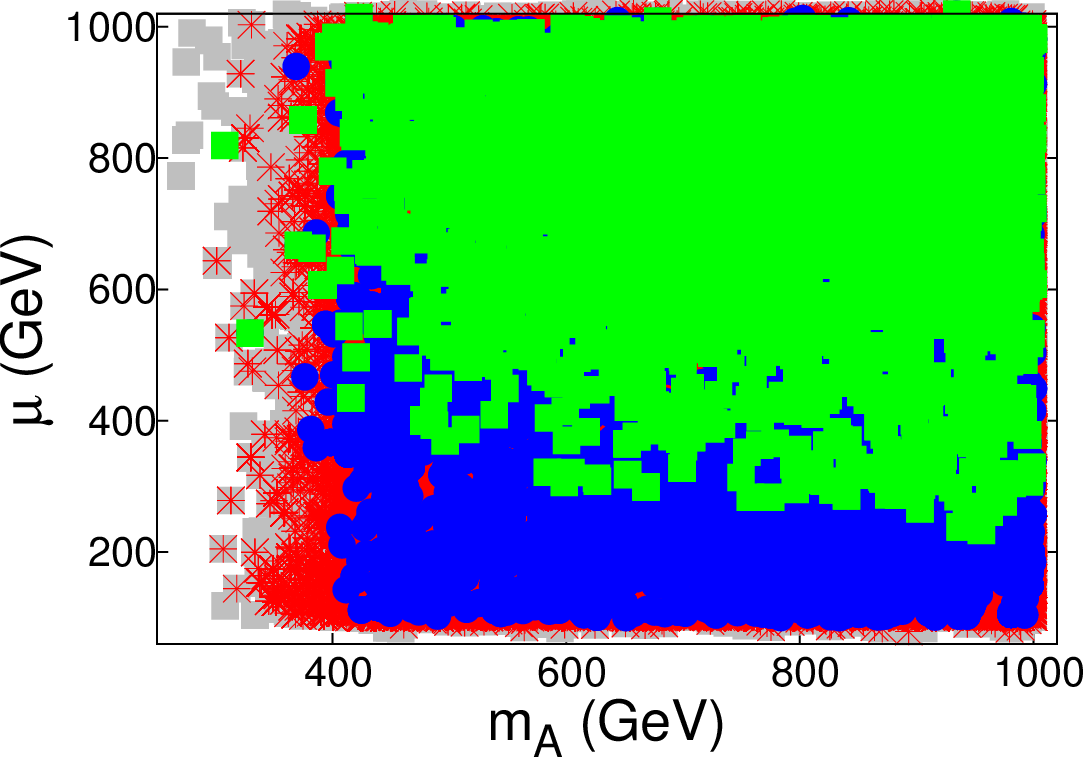}}
\subfigure[]{
      \includegraphics[width=209pt]{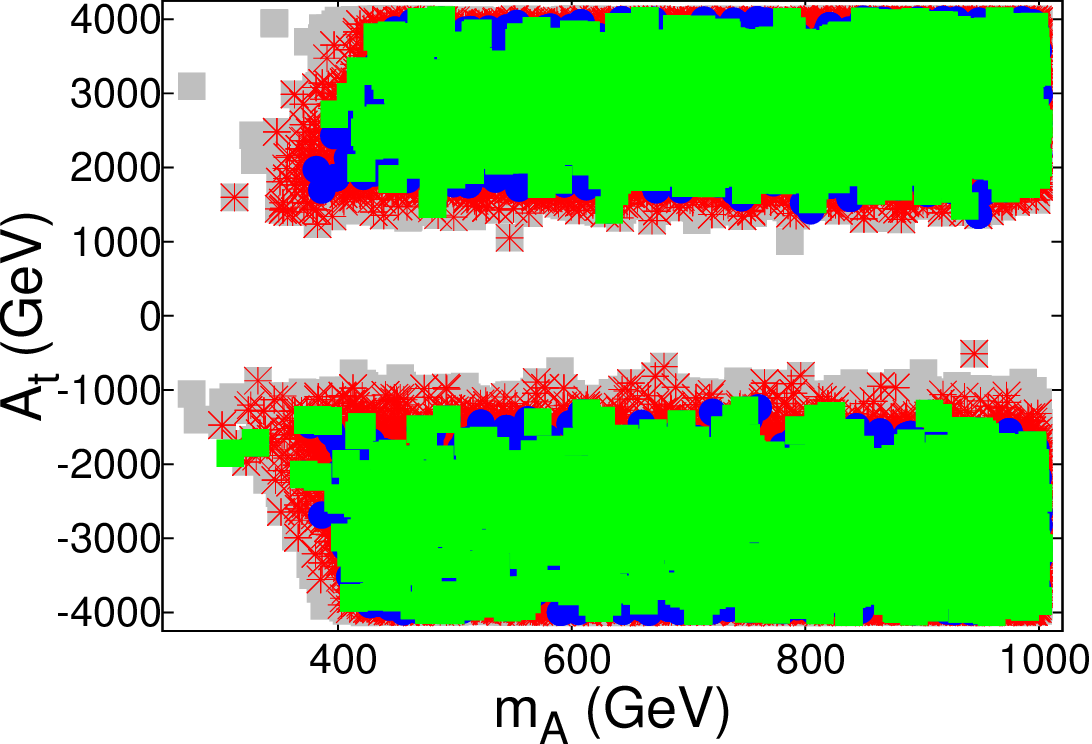}}
\subfigure[]{
      \includegraphics[width=209pt]{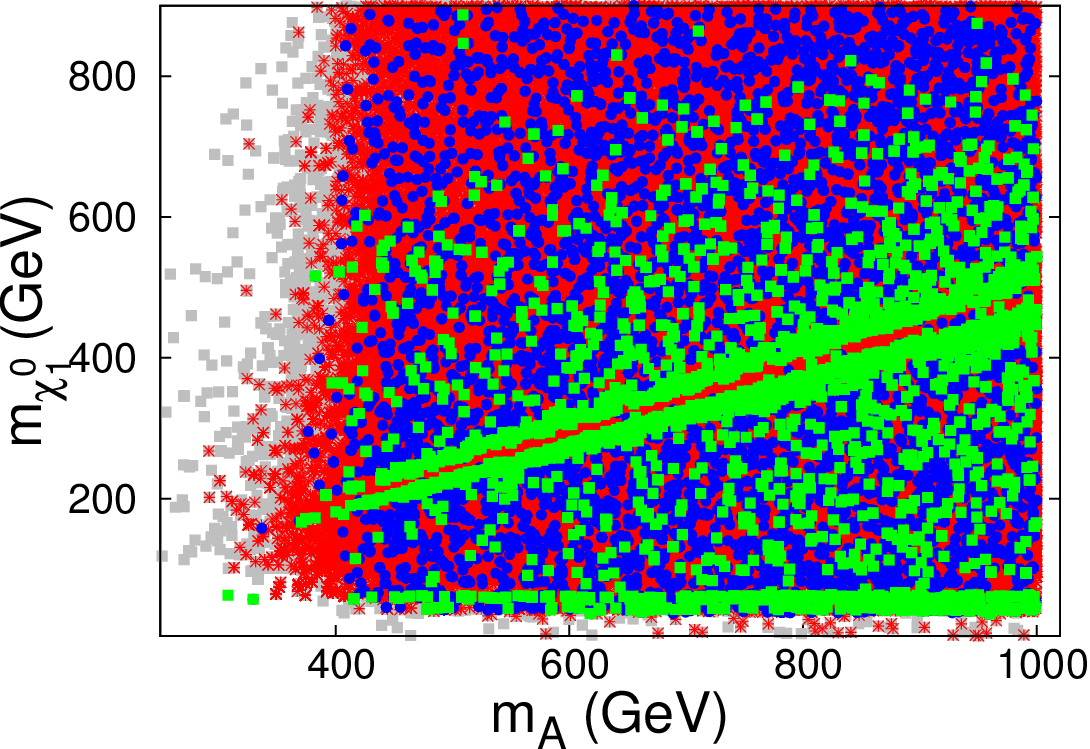}}
\end{center}
\caption[]{Allowed parameter regions versus the CP-odd Higgs boson mass $\ma$, for (a) $\tan\beta$,  (b) the Higgs mixing parameter $\mu$, (c) stop mixing parameter $A_{\rm t}$ and (d) LSP DM mass $m_\chi$, respectively.
%
%
All points pass the collider and Higgs constraints of eq.~(\ref{eq:requirement}). The grey squares require that the DM does not overclose the Universe; the red stars in addition satisfy the flavor constraints of eq.~(\ref{eq:flavor}); the blue disks are consistent with the LSP being all of the DM (i.e. predicts the correct relic density of eq.~(\ref{eq:relic_n})). The green squares pass the XENON-100 direct search bound in addition to the other requirements.}
\label{fig:para}
\end{figure}

We adopt the theoretical uncertainties from the SM predictions. We note that the uncertainties from experiments are of the same order of magnitude as the theoretical uncertainty for ${\rm BR}(b\to s\gamma)$, and thus the latter becomes very important.
In light of these precision results, we require our MSSM solutions to be within $2\sigma$ of the observed
value\footnote[2]{{It should be noted that the experimental measured value is an untagged value, while the theoretical prediction is CP averaged~\cite{DeBruyn:2012wj,DeBruyn:2012wk}.}}

\bea
\nonumber
2.79\times 10^{-4} < &{\rm BR}(b \to s\gamma) &<  4.07\times 10^{-4},\\
\label{eq:flavor}
1.1\times 10^{-9} < &{\rm BR}(B_{\rm s} \rightarrow \mu^+\mu^-) &< 5.3 \times 10^{-9}. \\
\nonumber
\eea

\begin{figure}[t]
\begin{center}
\subfigure[]{
\includegraphics[width=200pt]{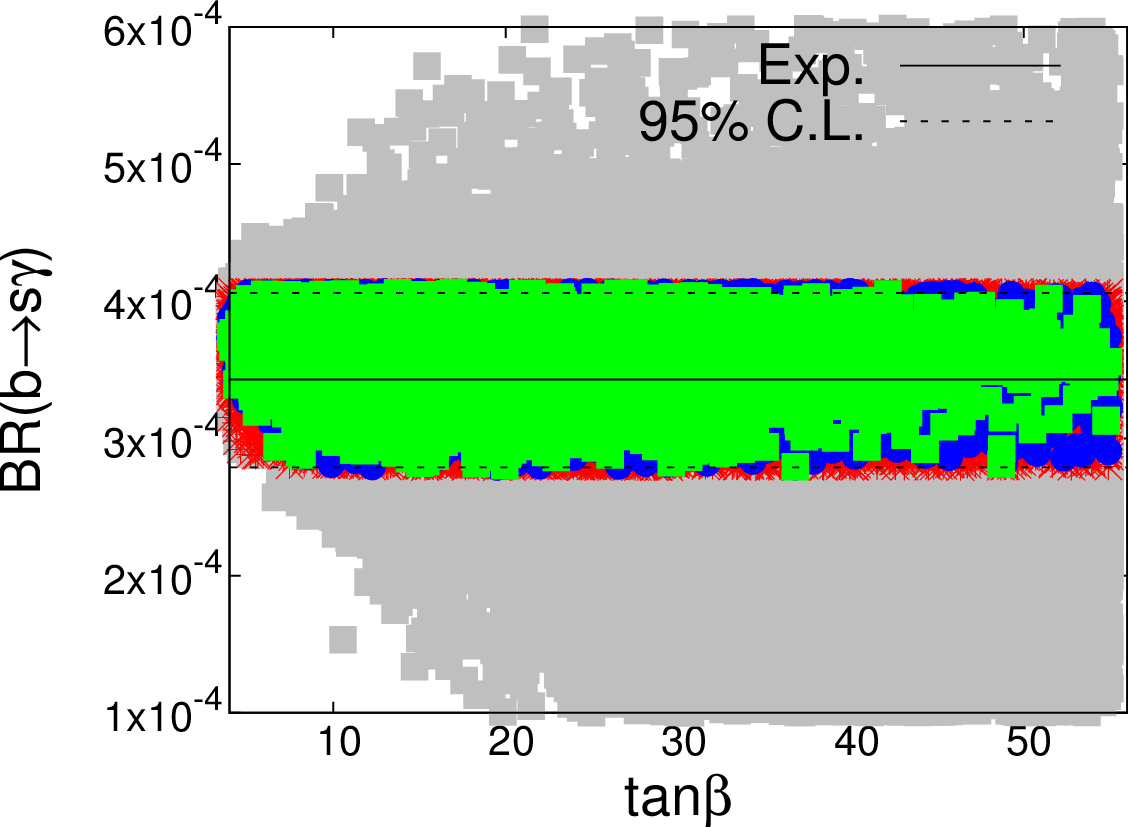}}
\subfigure[]{
\includegraphics[width=200pt]{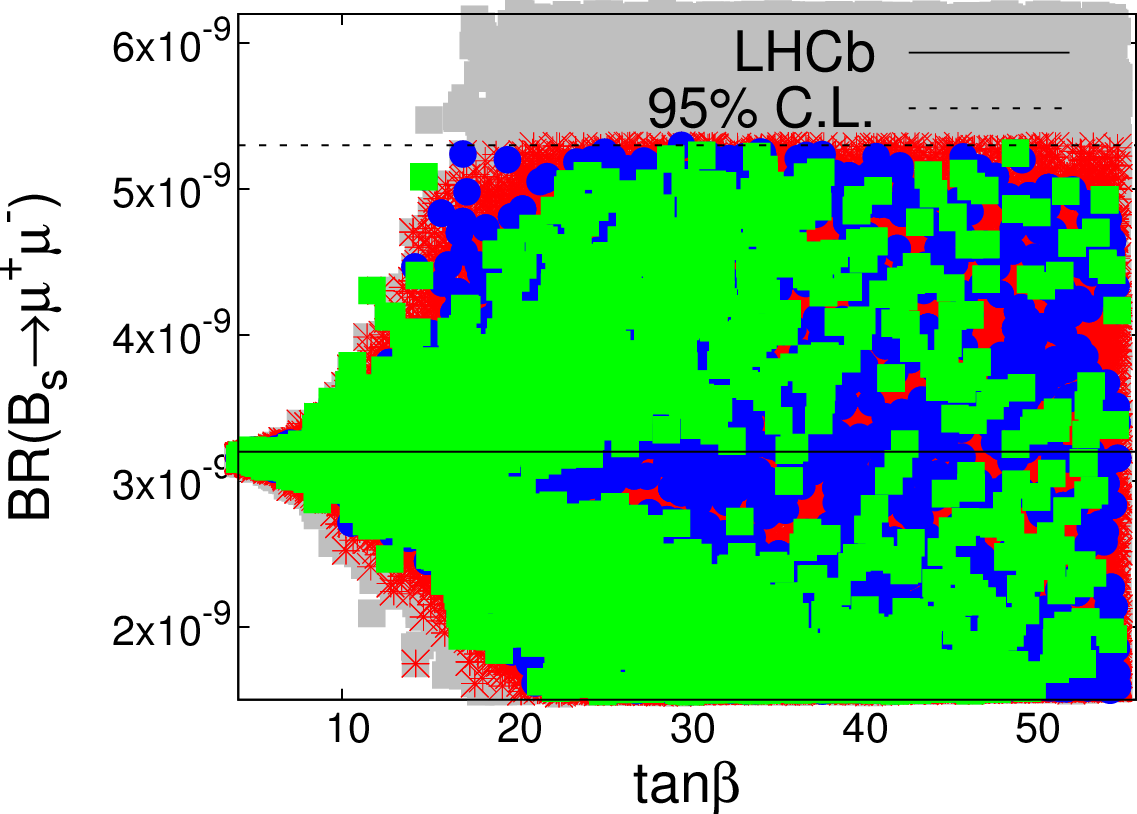}}
\end{center}
\caption[]{
Allowed branching fraction regions
versus $\tan\beta$, for (a) $b\to s\gamma$, (b)  $B_{\rm s} \rightarrow \mu^+\mu^-$.
The corresponding experimental central values and 2$\sigma$ bands are plotted on each panel.
Symbols and legends are the same as in figure~\ref{fig:para}.
\label{fig:flavor} }
\end{figure}


\subsection{Confronting the direct and indirect searches}

Thus far, the most stringent constraints on the spin-independent elastic scattering cross section ($\sigma^{\rm SI}_{\rm p}$) come from the XENON-100 experiment. The results from the XENON experiment challenge the signal hints from DAMA, CoGeNT and CRESST in the low mass region of $m_{\chi}\approx 10$ GeV, and cut deeply into the parameter space with $\sigma^{\rm SI}_{\rm p} \sim 2 \times 10^{-9}$ pb at $m_\chi \sim 60$ GeV. Limits on the spin-dependent cross section are not as constraining. We account for the bounds from the Super-Kamiokande \cite{superK}, and the IceCube/DeepCore \cite{icecube} experiments that are sensitive to the spin-dependent scattering of DM with Hydrogen at the sun's location. We also take into account bounds obtained by the Fermi satellite from the absence of gamma rays from the nearby dwarf galaxies.

\begin{figure}[htbp]
\begin{center}
\subfigure[]{
\includegraphics[width=208pt]{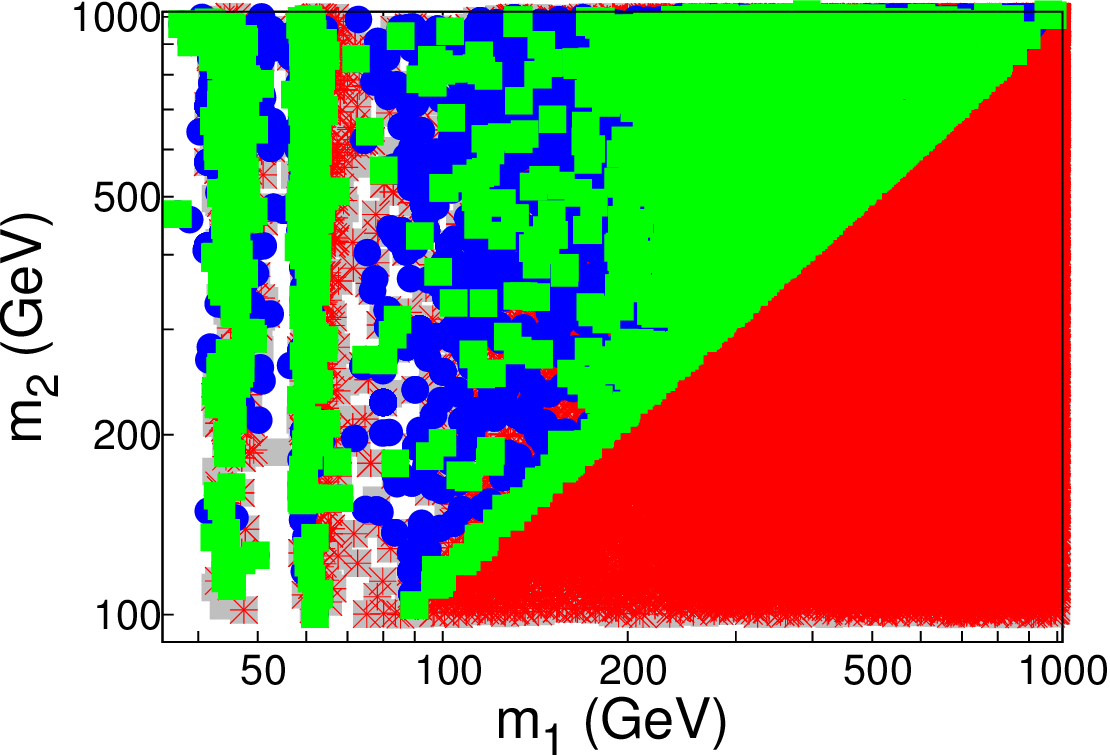}}
\subfigure[]{
\includegraphics[width=208pt]{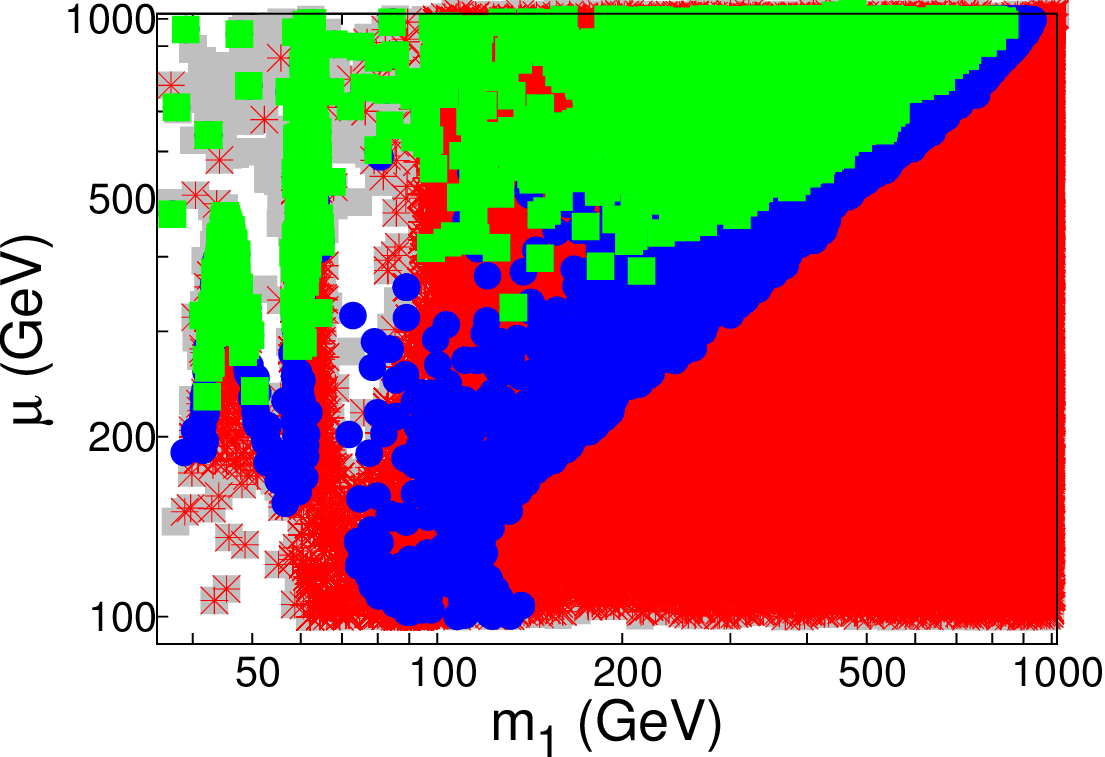}}
\subfigure[]{
\includegraphics[width=208pt]{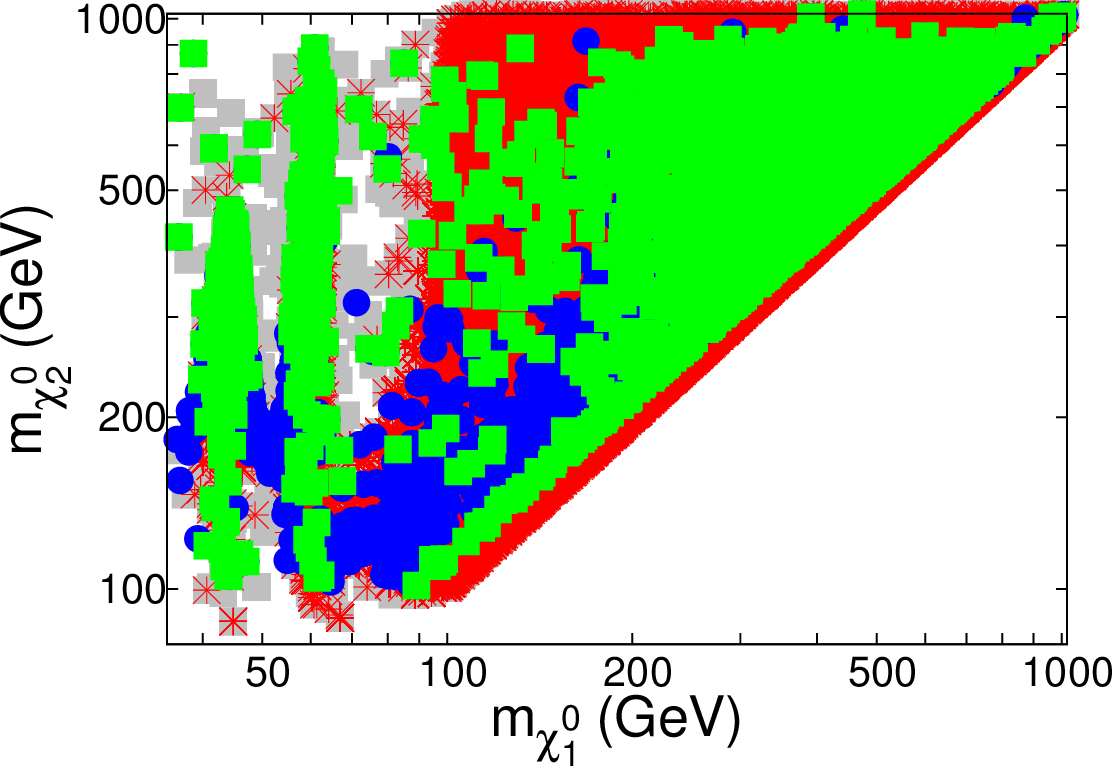}}
\subfigure[]{
\includegraphics[width=208pt]{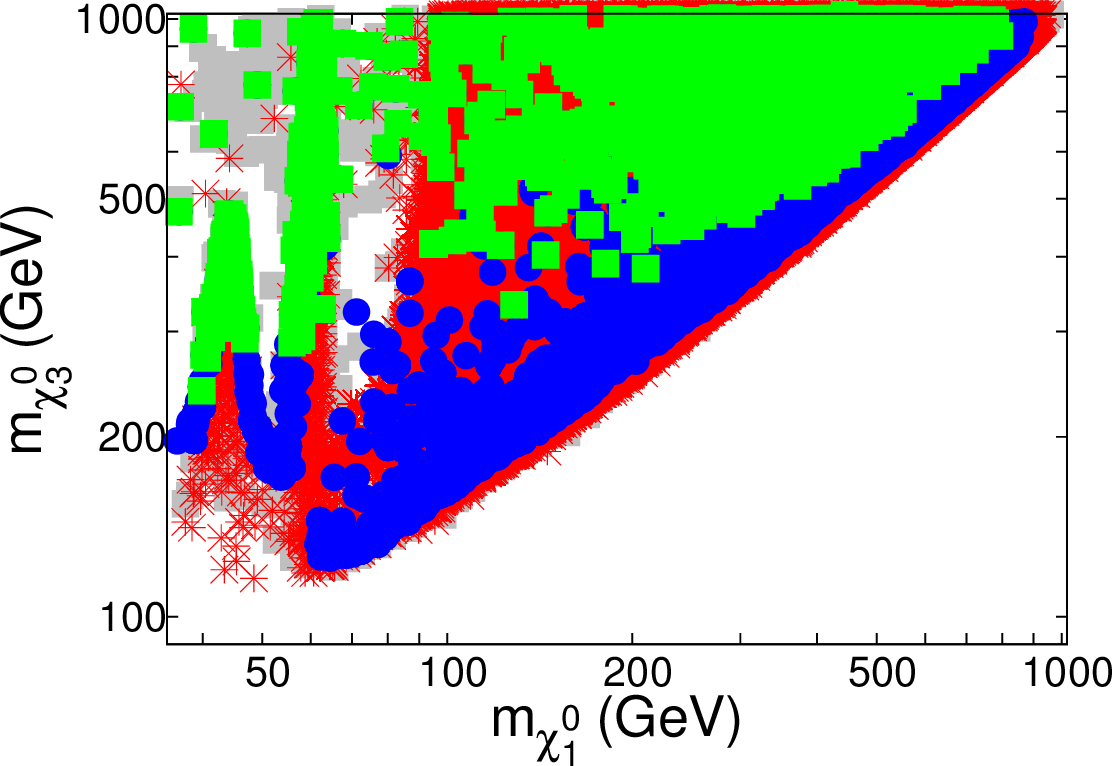}}
\subfigure[]{
\includegraphics[width=208pt]{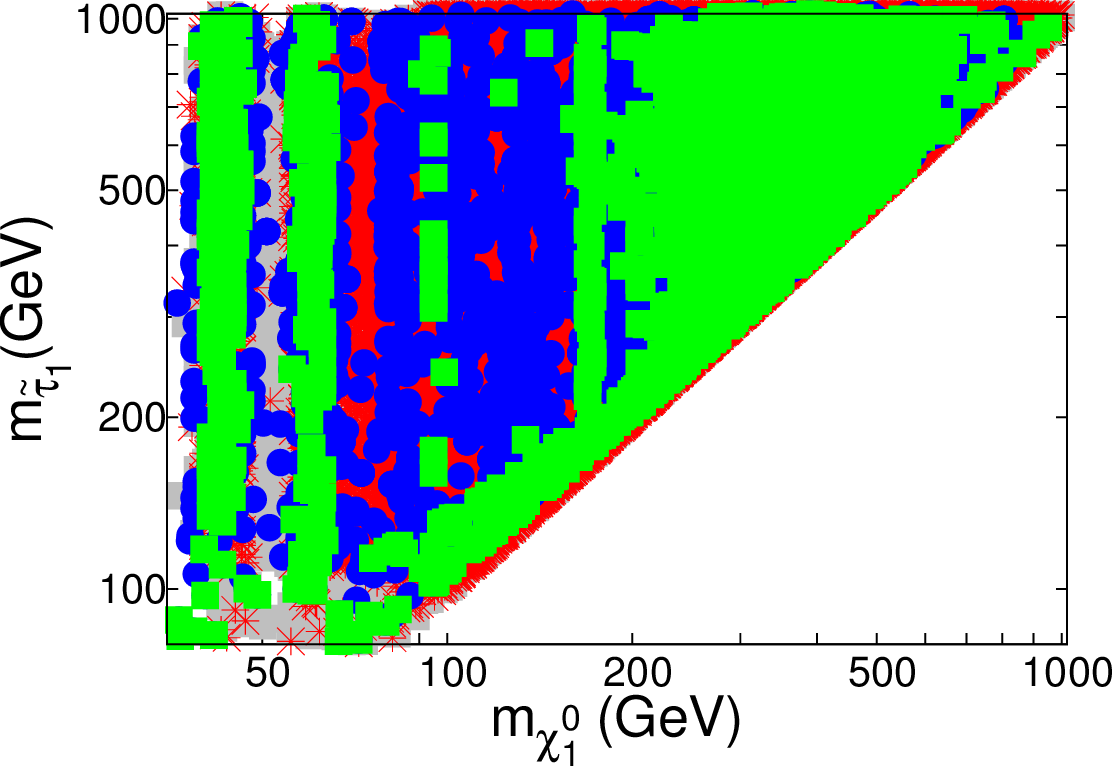}}
\subfigure[]{
\includegraphics[width=208pt]{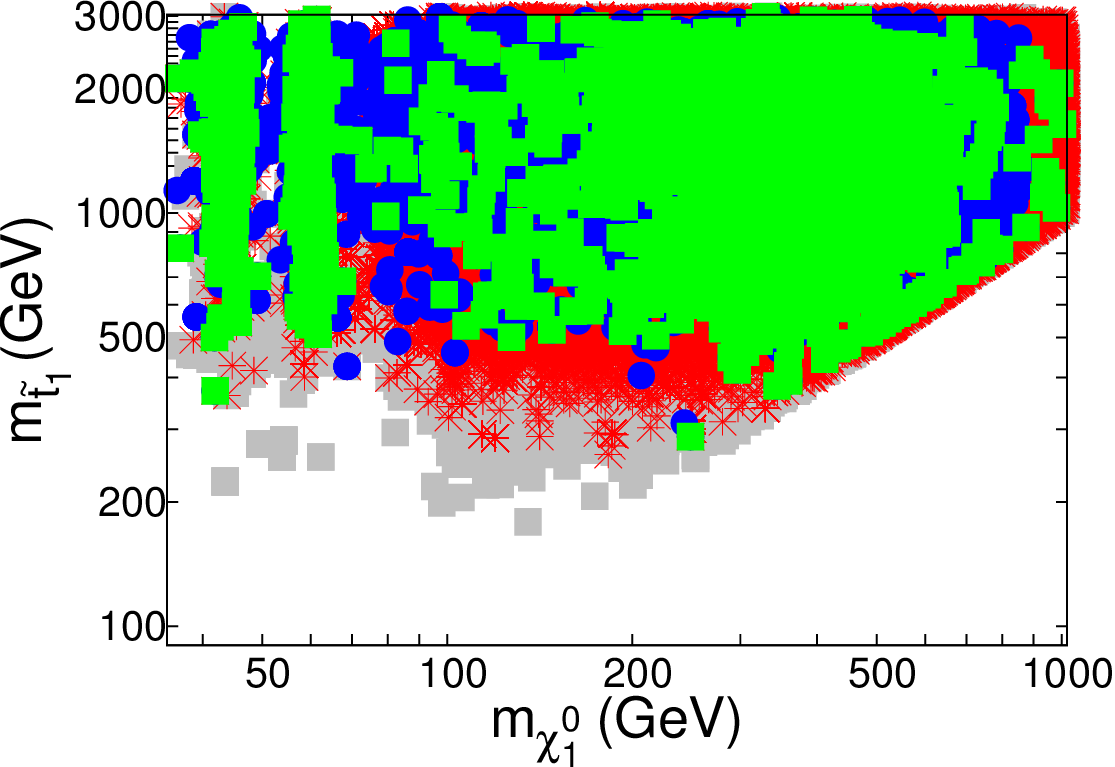}}
\end{center}
\caption[]{
Allowed parameter regions. Symbols and legends are the same as in figure~\ref{fig:para}.
(a) The Wino mass parameter $M_2$ and (b) the Higgsino mass parameter $\mu$ versus the Bino mass parameter $M_1$; (c) the second neutralino mass $m_{\chi_2^0}$ and (d) the third neutralino mass $m_{\chi_3^0}$ versus the lightest neutralino DM mass $m_{\chi_1^0}$;
(e) the lighter stau mass $m_{\tilde \tau_1}$ and (f) lighter stop mass $m_{\tilde t_1}$ versus the DM mass $m_{\chi_{1}^{0}}$.
}
\label{fig:para2}
\end{figure}


\subsection{Scanning results}
\label{results}

\begin{figure}[!t]
\begin{center}
\scalebox{0.375}{\includegraphics{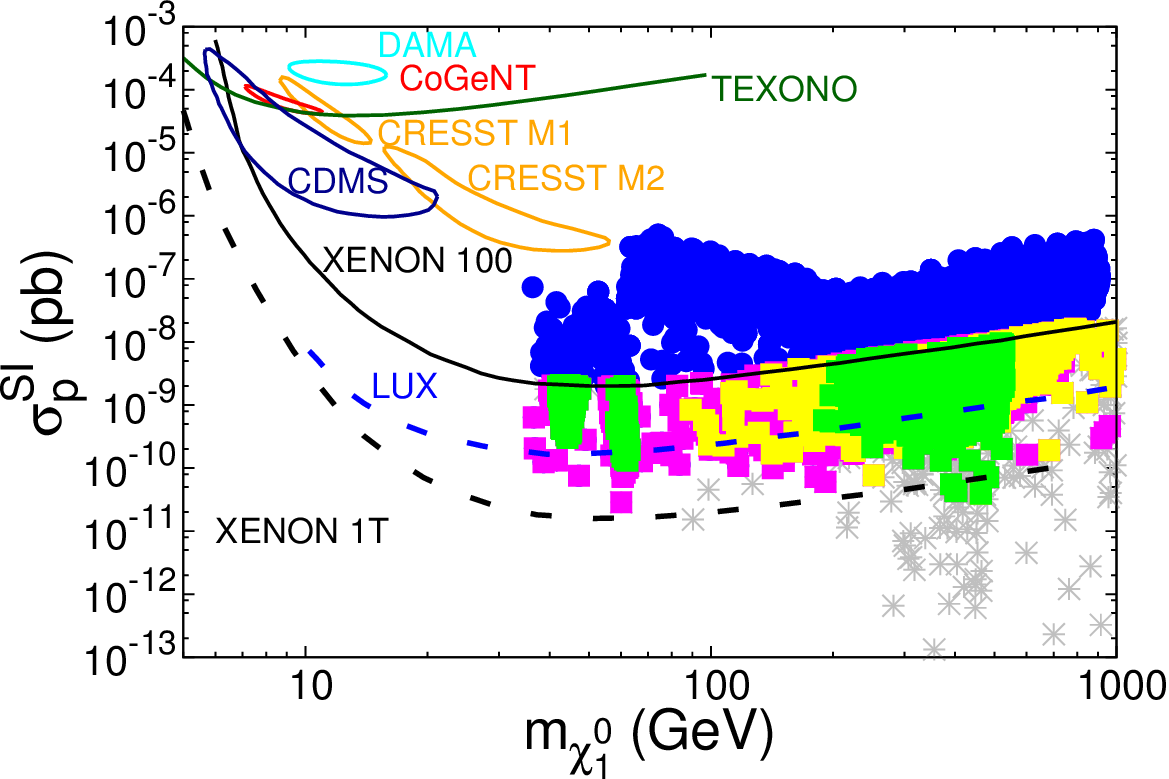}}
\end{center}
\caption[]{Spin-independent cross section versus the DM mass $m_{\chi_{1}^{0}}$. All the points in the colored shaded region give the correct relic abundance in eq.~(\ref{eq:relic_n}), satisfy the collider constraints in eq.~(\ref{eq:requirement}) and the flavor constraints in eq.~(\ref{eq:flavor}).
The green region represents the model points with the $Z$ and Higgs resonances. The $Z$ funnel and $h$ funnel regions are clearly visible for WIMP masses around half the $Z$ mass and half the Higgs mass.
The yellow points represent the region of co-annihilation with Wino-like/Higgsino-like NLSPs.
The magenta points represent the region with $\tilde \tau,~\tilde \nu_\tau,~\tilde b,~\tilde t$ contributions.
The gray points represent the scenarios with special cancellations when $M_{1}$ and $\mu$ take opposite signs.
The DAMA and CoGeNT  contours ($3\sigma$) are shown for astrophysical parameters $v_0$ = 220 km/s, $v_{\rm esc}$ = 600 km/s, and for a local density $\rho_0$ = 0.3 GeV/cm$^3$. CRESST contours are $2\sigma$ regions, from \cite{cresst}.
Also shown is the 90\% confidence contour from the recent CDMS analysis \cite{new_cdms}.
%
The blue region is excluded by the \xen experiment (90\% exclusion curve from \cite{xenon}, for $v_0$ = 220 km/s, $v_{\rm esc}$ = 544 km/s, $\rho_0$ = 0.3 GeV/cm$^3$).  Recent results from the TEXONO \cite{texono} collaboration are shown. Expected exclusion bounds from the ongoing LUX experiment \cite{lux} and the future XENON-1T experiment \cite{X1T} are also shown.
\label{fig:xenon}}
\end{figure}

\begin{figure}[tb]
\begin{center}
\subfigure[]{
\includegraphics[width=209pt]{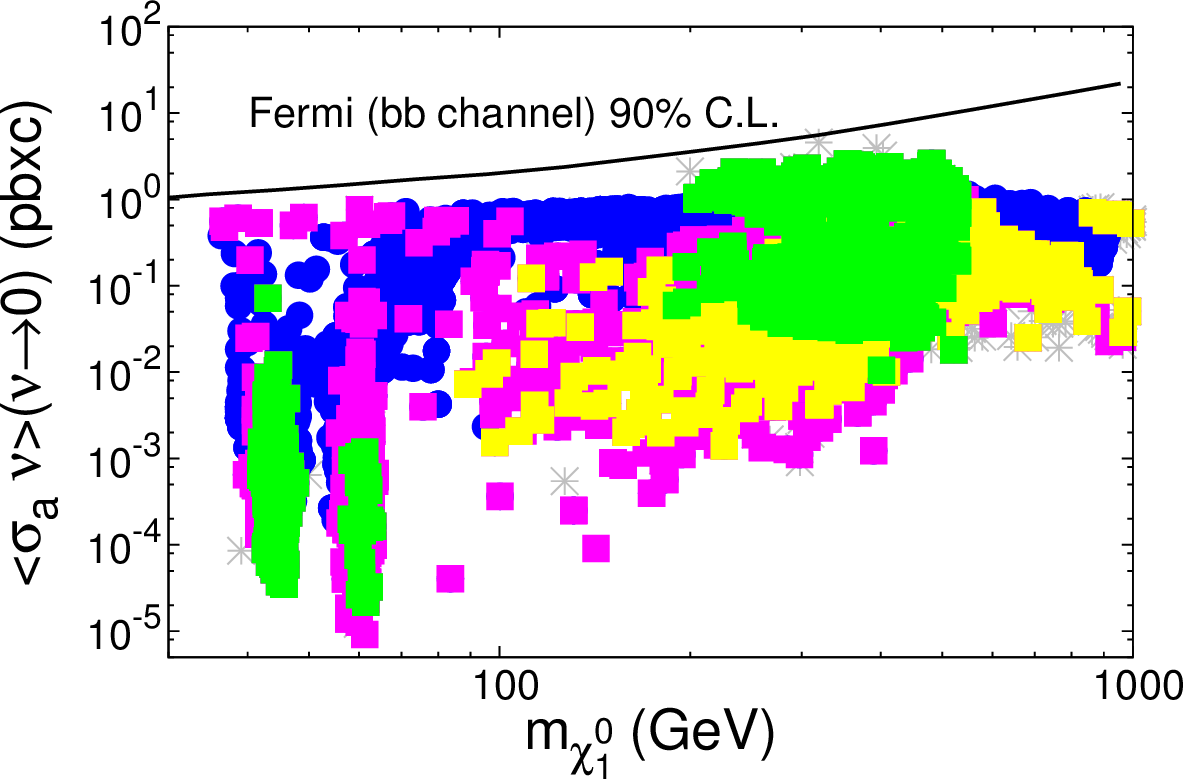}}
\subfigure[]{
\includegraphics[width=209pt]{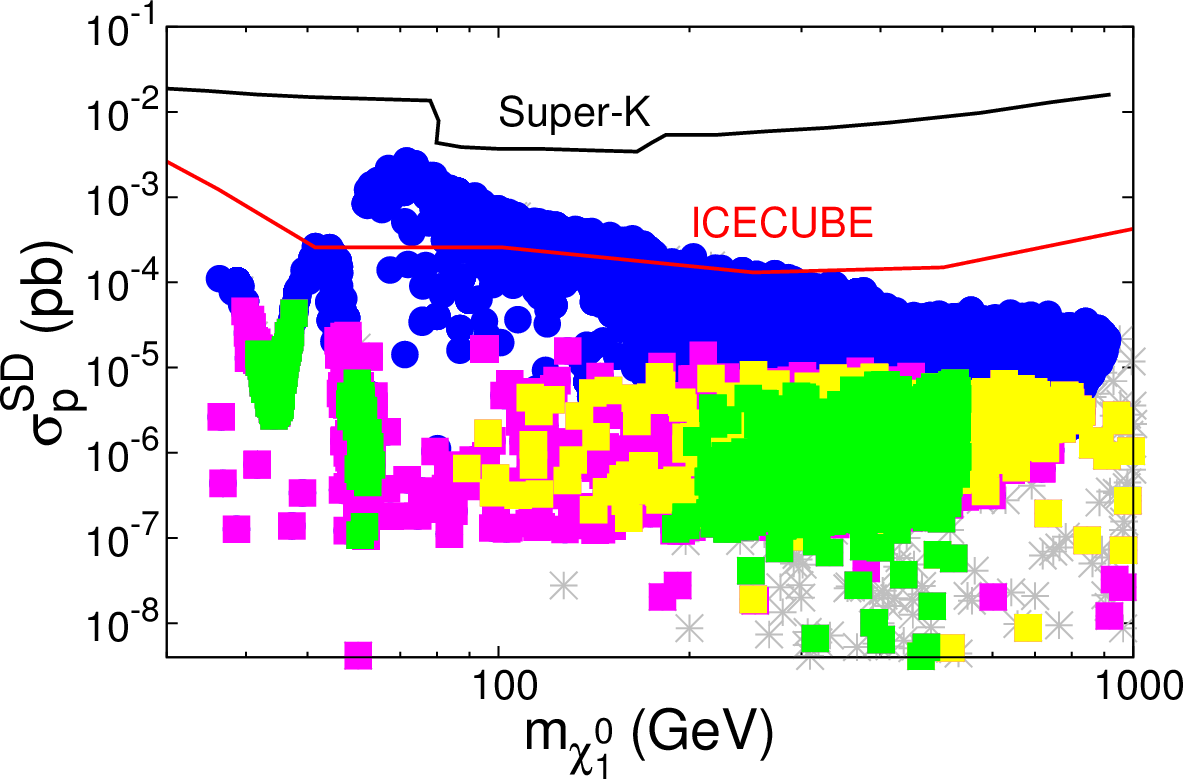}}
\end{center}
\caption[]{
(a) The annihilation cross section $\langle \sigma_{\rm a} v \rangle$ in the limit  $v \rightarrow 0$ along with the $95\%$ exclusion obtained by the Fermi satellite from the absence of gamma rays from the nearby dwarf galaxies \cite{fermi_dwarfs}.
(b) The spin-dependent scattering cross section with a proton, along with the 90\% exclusion curves from the Super-K \cite{superK} and IceCube \cite{icecube} experiments. Legends are the same as in figure~\ref{fig:xenon}.
}
\label{fig:cx}
\end{figure}

We now present our results for the allowed parameter regions in Figs.~\ref{fig:para}$-$\ref{fig:BRX}.
In figure~\ref{fig:para}, we show the parameter points passing the Higgs constraints in eq.~(\ref{eq:requirement}) versus the CP-odd Higgs boson mass $m_A$, for (a) $\tan\beta$,  (b) Higgs mixing parameter $\mu$, (c) stop mixing parameter $A_{\rm t}$ and (d) DM mass $m_{\chi_1^0}$, respectively.
These allowed parameter regions are shown in figure~\ref{fig:flavor} for branching fractions (a) BR($b\to s\gamma$) and (b)  BR($B_{\rm s} \rightarrow \mu^+\mu^-$) versus $\tan\beta$.
We show from the same set of points, the Wino mass parameter $M_2$ and the Higgsino mass parameter $\mu$ versus the Bino mass parameter $M_{1}$ in Figs.~\ref{fig:para2}(a) and (b). We show the second and third neutralino masses $m_{\chi_2^0}$, $m_{\chi_3^0}$, the light stau mass and the light stop mass versus the LSP mass $m_{\chi}$ in Figs.~\ref{fig:para2}(c), (d), (e) and (f).
In the above Figures \ref{fig:para}$-$\ref{fig:para2}, all points satisfy the collider, and Higgs search requirements in eq.~\ref{eq:requirement}. The grey squares show MSSM models that do not overclose the universe. The red stars in addition satisfy the flavor requirements in eq.~(\ref{eq:flavor}). The blue disks represent the models that give the correct relic density in eq.~(\ref{eq:relic_n}). Finally, the green squares pass the severe XENON-100 direct search bound on the WIMP-proton spin-independent elastic scattering.

The results obtained here are consistent with the existing literature on the studies at the LHC \cite{tao_higgs_paper,Carena:2013qia}. We make the following important observations: \\

\noindent
\underline{(1). Higgs constraints (grey squares):}
We start with points that do not overclose the universe and satisfy the collider search requirements in eq.~(\ref{eq:requirement}). We reproduced the known results that there are two surviving regions:

\noindent
(i) The non-decoupling regime where $m_{\rm A} \sim 95-130$ GeV, the heavy CP-even Higgs $(H)$ is SM-like, and the light CP-even Higgs ($h$) is nearly degenerate in mass with the CP-odd Higgs ($A$). This region is particularly interesting since it leads to rich collider phenomenology and favors a light WIMP mass $m_\chi \lesssim 50$ GeV. These points are not shown on the plots since they are disfavored by the flavor constraints, as discussed next. \\

\noindent
(ii) The decoupling regime where $m_{\rm A} \gtrsim 250$ GeV, the light CP-even Higgs is SM-like, and the heavy CP-even Higgs is nearly degenerate in mass with the CP-odd Higgs. This regime is difficult to observe at the LHC when $m_{\rm A} \gtrsim 400~\gev$ and $\tan\beta\sim 10-20$  in traditional SM Higgs search channels due to severely suppressed couplings to the gauge bosons. \\

\noindent
\underline{(2). Flavor constraints (red stars):}
The two decay processes $b\to s\gamma$ and $B_{\rm s} \rightarrow \mu^+\mu^-$ are the most constraining ones.
The experimental central values are plotted on the calculated branching fractions in figure~\ref{fig:flavor} on each panel, along with 2$\sigma$ bands, which is summarized in eq.~(\ref{eq:flavor}). These flavor constraints prefer lower $\tan\beta$ values and essentially remove the light Higgs ($H^{0},A^{0},H^{\pm}$) solutions in the non-decoupling region in our generic scan.
The solutions with a light LSP of $m_{\chi} \lesssim 30$ GeV are also eliminated. Our results are in good agreement with the existing studies \cite{arbey_etal_1, arbey_etal_2}. Further improvements in the $B_{\rm s} \rightarrow \mu^+\mu^-$ measurement would strongly constrain the large $\tan\beta$ and low $\ma$ region.
However, we have not tried to exhaust parameter choices with possible cancellations among different SUSY contributions, and some sophisticated scanning may still find solutions with certain degrees of fine-tuning \cite{Tongetal}. \\

\noindent
\underline{(3). Relic density requirement (blue disks):}
Merely requiring that the LSP does not overclose the universe does not constrain the MSSM parameter space very much, as most clearly seen from the gray squares and red stars in figure~\ref{fig:para2}. This is because the Higgsino-like or Wino-like LSPs and NLSPs can annihilate efficiently through gauge bosons and Higgs bosons. Requiring the correct dark matter relic density at the present epoch does constrain the parameter space significantly. We see the preference for $\mu>M_{1}$ and $M_{2}>M_{1}$, as in Figs.~\ref{fig:para2}(a) and (b). Otherwise the Higgsino or Wino LSP would annihilate too efficiently, and result in underabundant DM relic.
Nevertheless, we do find a nearly degenerate region of a Bino LSP and Wino NLSPs as seen in figure~\ref{fig:para2}(c), which is best characterized by the ``well-tempered'' scenario \cite{ArkaniHamed:2006mb}. This scenario, however, seems to be less implementable with Higgsino NLSPs as seen in Figs.~\ref{fig:para2}(b) and (d), if $\mu$ is not much greater 1 TeV.
Importantly for our interests, we see prominent strips near $m_{\chi} \sim m_{Z}/2,\ m_{h}/2$ which are the $Z$ and Higgs funnel regions. Interestingly, as seen in figure~\ref{fig:para}(d), there is a region of depletion near $m_{A}\approx 2 m_{\chi}$, indicating the very (too) efficient annihilation near the $A^{0}$ funnel in the $s$-channel that is removed by the correct relic density requirement. This is a result of a lower bound on the LSP-Higgsino component $N_{13}$ that we will discuss later.

For the low mass dark matter that is favored by CoGent, DAMA, CRESST and CDMS experiments, the solutions are disfavored by precision electroweak observables, LEP constraints on SUSY direct searches, and constraints from the Higgs property\footnote[3]{Our requirement of the $h\to \gamma\gamma$ rate in eq.(\ref{eq:requirement}) also limits the allowed Higgs branching fractions to SUSY particle pairs, especially for those solutions with kinematically allowed Higgs decays to NLSP pairs.}.
In our analysis, we strictly apply the LEP bounds on the SUSY searches and the requirement for a SM-like Higgs boson as given in eq.~(\ref{eq:requirement}), then there are no surviving points in the low mass DM region.
However, as noted in Refs.~\cite{Arbey:2012na,Boehm:2013qva,Arbey:2013aba,Belanger:2013pna,Hagiwara:2013qya}, if one adopt the scenarios with a compressed spectrum, such as a mass difference $m_{\tilde b}-m_{\chi}<5~\gev$  to evade the LEP bounds, or relax the $h\to \gamma\gamma$ to be SM-like, new solutions in the low mass region could emerge.

\noindent
\underline{(4). Direct search bounds (green,  yellow and  magenta squares):} The results from DM direct searches can be translated to spin-independent cross sections and thus to the MSSM parameters. This is shown in figure~\ref{fig:xenon}, where all the points in the colored shaded region give the correct relic abundance in eq.~(\ref{eq:relic_n}), satisfy the collider constraints in eq.~(\ref{eq:requirement}) and the flavor constraints in eq.~(\ref{eq:flavor}).
The parameter space favored by the DAMA, CoGeNT, CRESST and CDMS experiments, as well as the stringent bound from the XENON-100 experiment are plotted. We see that the blue region is further excluded by the \xen experiment\footnotemark[4]. As seen in figure~\ref{fig:para}(a), lower $\tan\beta$ and higher $\ma$ values are preferred. Figures \ref{fig:para}(b) and \ref{fig:para2}(b) show the lower bound $\mu>$ 200 GeV. This consequently leads to a heavier $\chi^{0}_{3}$ as seen in \ref{fig:para2}(d), while $\chi^{0}_{2}$ could be still as light as the LSP $\chi^{0}_{1}$ as seen in \ref{fig:para2}(c).
\footnotetext[4]{It should be noted that the theoretical calculation of the spin-independent cross section may have significant uncertainties~\cite{Ellis:2008hf,Accomando:1999eg}.}

The most important observation from our study is that the surviving points are quite characteristic.
We can identify the following classes of predictive features for the LSP DM from figure~\ref{fig:xenon}.

\begin{itemize}
\item[I-A]
(green)~~~~$\chi_{1}^{0}\chi_{1}^{0} \to Z \to SM$ predicts $m_{\chi} \approx \mz/2 \sim 45$ GeV, the $Z$-funnel~\cite{Belanger:2001am}.
\item[I-B]
(green)~~~~$\chi_{1}^{0}\chi_{1}^{0} \to h \to SM$ predicts $m_{\chi} \approx m_{h}/2 \sim 63$ GeV,  the $h$-funnel.
\item[I-C]
(green)~~~~$\chi_{1}^{0}\chi_{1}^{0} \to H,A \to SM$ predicts $m_{\chi} \approx m_{A,H}/2 \sim 0.2-0.5$ TeV, the $H/A$-funnel. The A-funnel is overall dominant comparing to the H-funnel.
\item[II-A]
(yellow)~~~Neutralino/chargino coannihilation~\cite{coan2,coan5}: $\chi^{0}_{i} \chi^{0}_{j},\ \chi^{0}_{i} \chi^{\pm}_{j}\to SM.$
\item[II-B]
(magenta)~Sfermion assistance~\cite{coan1,coan3,coan4}: $\chi_{1}^{0}\tilde{\tau},~\chi_{1}^{0}\tilde{t},~\chi_1^0\tilde b \to SM$;
$t$-channel $\tilde\tau,\ \tilde \nu$ in $\chi^{0}_{i} \chi^{0}_{j}$.
\end{itemize}

We categorize model points as scenario I if the difference between the mediator mass and twice the LSP mass is within $8\%$ of the mediator mass, namely
\beq
|m_{Z,h,A}-2m_{\chi^0_1}| \le 0.08~m_{Z,h,A}.
\eeq


We categorize model points as scenario II-A if the difference between the LSP mass and neutralino NLSP\footnotemark[5] mass is less than $15\%$ of the LSP mass, namely $m_{\chi_2^0}-m_{\chi_1^0}<0.15m_{\chi_1^0}$. Other cases are categorized as scenario II-B. Our classification and categorization have been verified by investigating a fraction of our generated model points and looking into their individual contributing annihilation channels. Two remarks are in order: First, the light sfermions needed in this category are still viable, especially for $\tilde t,\ \tilde b$, with respect to the direct LHC searches, because the mass splitting with the LSP is too small to result in significant missing transverse energy to be sensitive for the search. In case of compressed spectra, LHC searches on the monojet and monophoton could gain some sensitivity~\cite{Carena:2008mj,Rajaraman:2011wf,Fox:2011pm,Delgado:2012eu,Lin:2013sca,An:2013xka}. Second, this categorization based on simple kinematics has exemptions when the LSP coupling to the resonant mediator is very week and thus the co-annihilation mechanism is dominant. We have kept track of those cases in the plot by examining the points individually. \\
\footnotetext[5]{This is almost always true because we have a very Bino-like LSP. For cases with $\tilde \tau_1, ~\tilde t_1$ NLSP with the sfermion coannihilation mechanism, they fall into scenario II-B automatically.}

\noindent
\underline{(5). Indirect search bounds:}

There exist cosmological bounds from the indirect search for DM signals. We present the annihilation cross section $\langle \sigma_{\rm a} v \rangle$ in the limit  $v \rightarrow 0$ (i.e. the $v$-independent component) versus the LSP DM mass in figure~\ref{fig:cx}(a), along with the 95\% exclusion obtained by the Fermi-LAT satellite from the absence of gamma rays from the nearby dwarf galaxies \cite{fermi_dwarfs}. We see that further improvement from the measurement by the Fermi-LAT will reach the current sensitivity range.
The spin-dependent scattering cross section with a proton is shown in figure~\ref{fig:cx}(b), along with the 90\% exclusion curves from the Super-Kamiokande experiment \cite{superK} and the IceCube constraint from DM annihilation in the Sun~\cite{icecube}. We see that IceCube results are cutting into the relevant parameter region closing the gap from the direct searches, although the bounds from the indirect searches are not quite as strong as that from XENON-100. It is worth mentioning that the local DM density in the Sun may be higher than the referral value~\cite{Salucci:2010qr} and thus would yield a slightly stronger exclusion from IceCube.


\section{Discussions}
\label{Discuss}

\subsection{The nature of the DM}

\begin{figure}[t]
\centering
\subfigure[]{
\includegraphics[width=209pt]{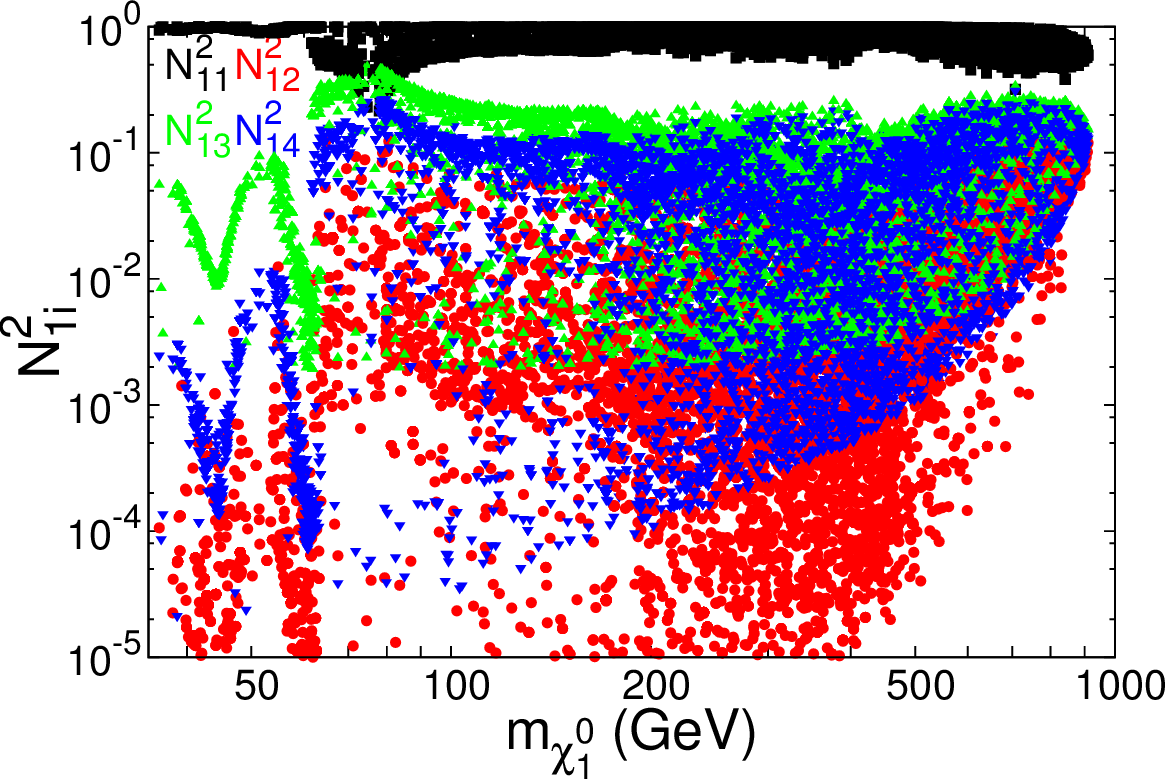}}
\subfigure[]{
\includegraphics[width=209pt]{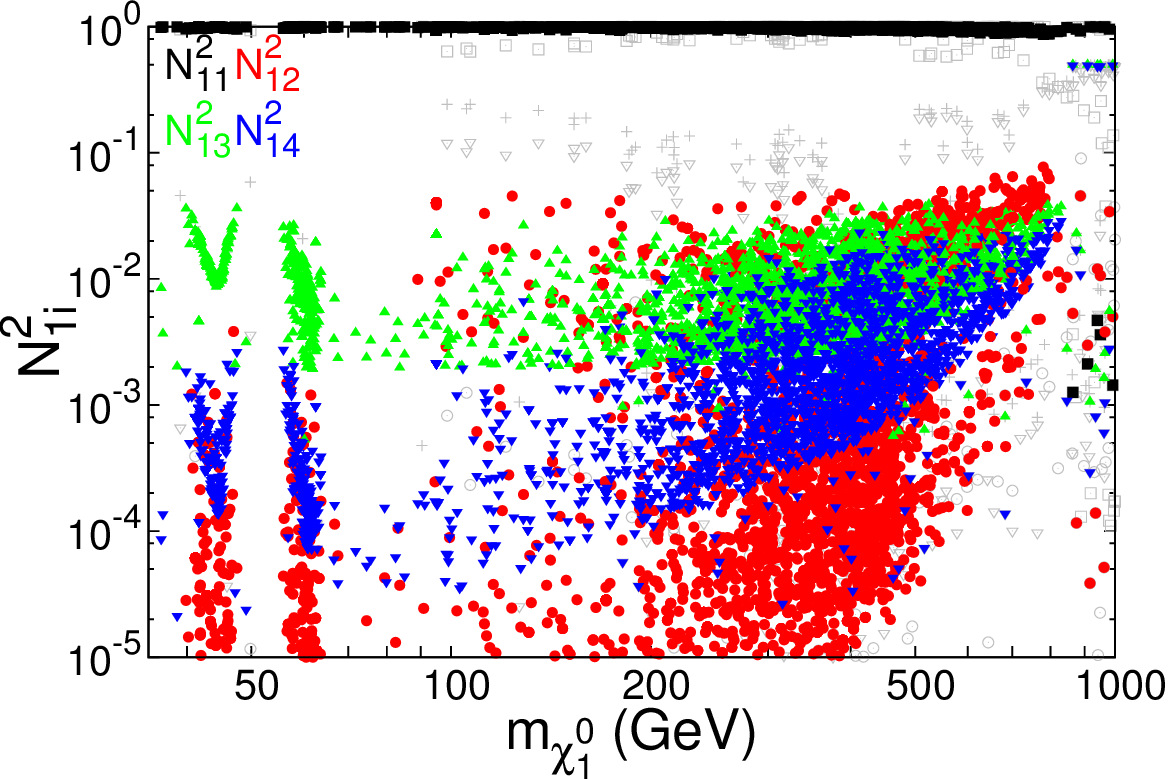}}
\caption[]{The gaugino and Higgsino fractions of the LSP versus $m_{\chi^{0}_{1}}$ (a) consistent with the relic density, collider, and flavor constraints, (b) consistent with XENON-100 in addition to the other requirements.
The gray points represent the results for $M_1$ and $\mu$ to have opposite signs and the corresponding fractions $N_{11}^2$, $N_{12}^2$, $N_{13}^2$ and $N_{14}^2$ are denoted by hollow squares, circles, daggers and hollow triangles, respectively.
\label{fig:fractions} }
\end{figure}

Experimental results from the collider searches, the $b$-quark rare decays and the direct DM searches, combined with the relic density requirement have put very stringent constraints on the SUSY parameter space. This in turn could have significant implications for searches at future collider experiments. Of primary importance is the nature of the LSP. We show the gaugino and Higgsino fraction ($N^2_{1i}$) of the neutralino LSP versus its mass in figure~\ref{fig:fractions}, consistent with all collider and flavor measurements as well as the correct relic density. From figure~\ref{fig:fractions}(a), we note that the surviving points are mostly Bino-like ($N^2_{11}$, as shown by the black dots), with lower fractions for Wino-like ($N^2_{12}$, red dots) and Higgsino-like ($N^2_{13},\ N^2_{14}$, green and blue dots, respectively). As noted earlier, this is because Wino-like and Higgsino-like LSP's annihilate very efficiently via SU(2) gauge interactions resulting in too little dark matter at the present epoch. Yet, the LSP could not be purely Bino-like since it would overclose the Universe.
In the region $m_{\chi^{0}_{1}} \sim $ 40 GeV$-$60 GeV, the line structures corresponding to the Higgsino components are easily identifiable with the $Z$ and $h$ exchanges.

The \xen direct search plays a crucial role in fixing the DM properties. The surviving points are shown in figure~\ref{fig:fractions}(b). We see that the Wino and Higgsino fractions of the LSP are further constrained. The $\tilde H_{d}$ component comes in with the opposite signs with respect to the $\tilde H_{u}$ and $\tilde W$ components. Bino-like LSP becomes more pronounced and the Wino and Higgsino  components consist of less than $7\%$ each,
rendering the ``well-tempered'' scenario \cite{ArkaniHamed:2006mb} difficult to realize with large Bino-Wino or Bino-Higgsino mixing.
The comparison between figure~\ref{fig:fractions}(a) and (b) clearly shows the XENON-100 exclusion probes deeply into the Higgsino and Wino components. On the other hand, the $\tilde H_{d}$ component $N^2_{13}$ must be non-zero, and so is $N^2_{14}$ for $\tilde H_{u}$. The non-zero lower bound would have significant implications for direct searches as we will discuss next, although the precise values may depend on the choice of the ranges for $M_{1},\ M_{2}$ and $\mu$.

It is important to note that a relative opposite sign between $M_{1}$ and $\mu$ could result in a subtle cancellation for the
$h\chi\chi$ coupling \cite{Perelstein:2011tg,Cheung:2012qy}, and thus enlarge the allowed mixing parameters, reaching about $20\%$ Wino/Higgsino mixtures,
as shown by the grey points in figure~\ref{fig:fractions}(b). This can happen only for a higher LSP mass when co-annihilations or $H,A$ funnels are in effect.

\subsection{Lower limit on the spin-independent cross section}

With our assumptions in the MSSM framework and the well-constrained properties of the LSP, we would expect that the DM scattering cross section may be predicted.It is interesting to ask whether one may derive a lower limit for the spin-independent scattering cross section. This is quite achievable for the Higgs resonance situation.
Much effort has been made to derive the neutralino recoil cross sections with nuclei in various SUSY models \cite{Neutralino3, srednicki_watkins,gelmini_gondolo_roulet,drees,drees_nojiri_3,drees_nojiri_2,ellis_ferstl_olive}.  This cross section mainly receives contributions from $h,\ H$ exchanges and squark exchanges. Given the current experimental bounds on the masses of the squarks from the LHC \cite{CMS-PAS-SUS-12-023,Aad:2012naa}, the Higgs exchanges dominate.
As a good approximation in the decoupling limit $\cos(\alpha-\beta)\simeq 0, \ \tan\beta\geq~3$ and a Bino-like LSP, the neutralino-nucleon cross sections via the Higgs exchanges are of the form \cite{drees_nojiri_2}:
\bea
\label{eq:hh}
\sigma_{\chi N}
& \propto &
\left\{
\begin{array}{ll}
\frac {|N_{11}|^2 |N_{13}|^2} {m^4_H \cos^2 \beta} (f_{T_s}+\frac 2 {27} f_{TG})^2,  & \quad  H \  {\rm  exchange}, \\
\frac {|N_{11}|^2 |N_{14}|^2} {m^4_h} (f_{T_u}+\frac 4 {27} f_{TG})^2, & \quad h\ {\rm exchange}.
\end{array}
\right.
\eea
$f_{T_s},~f_{T_u}$ and $f_{TG}$ are parameters measured from nuclear physics experiments. The $H$ exchange benefits from an enhancement factor $(N_{13}/\cos\beta)^{2}$.
When the $H$ is heavy, the $h$ exchange proportional to $N^2_{14}$ becomes important.
Due to our natural choices of parameters as in eq.~(\ref{eq:scan}), there exist lower bounds on $N^2_{13}$ and $N^2_{14}$, as shown in figure~\ref{fig:fractions}, although unnaturally large values of $\mu$ and $m_A$ could relax these bounds.  Consequently, the LSP spin-independent cross sections at the $Z,h$ funnels, which are mainly from the LSP scattering of a $t$-channel $H$ exchange, reaches a lower bound, roughly
\beq
\sigma^{SI}_{\rm p}(h,H) \gtrsim  10^{-10}\ {\rm pb.}
\label{eq:bound}
\eeq
As seen in figure~\ref{fig:xenon}, this range (green dots) will be probed by the ongoing LUX experiment and will be fully covered by the proposed  XENON-1T experiment.
Similar argument could be also applicable to the $H,A$ funnel regions, where $t$-channel $h$ exchange could become dominant. However, an exception is that a subtle cancellation takes place when $M_{1}$ and $\mu$ take opposite signs  \cite{Perelstein:2011tg,Grothaus:2012js,Cheung:2012qy}. As seen from the grey points in figure~\ref{fig:xenon}(b), this can happen only for a higher LSP mass when co-annihilations or $H,A$ funnels could be in effect.

In Ref.~\cite{leszek},
a parameter-independent lower bound $\sigma_{\rm p}^{SI} \gtrsim 2 \times 10^{-12}$ pb could be obtained in the mass range $440~\gev~\lesssim m_\chi \lesssim~1020~\gev$ and $\mu>0$. In the most general pMSSM \cite{CahillRowley:2012kx} with much larger $M_{2}, \mu$ parameters, the spin-independent cross section could go lower, depending on the
mixing parameters.
%

\subsection{Connection to the indirect searches}


\begin{figure}
\centering
\subfigure[]{
\includegraphics[width=205pt]{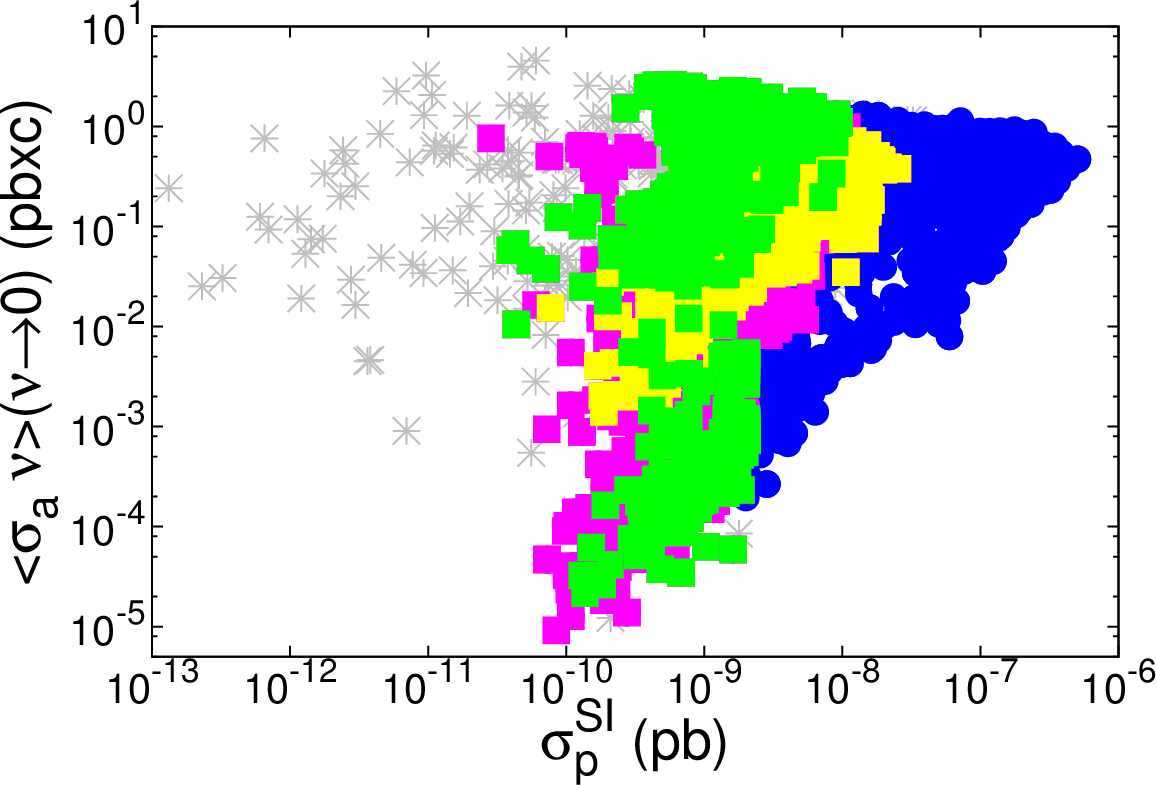}}
\subfigure[]{
\includegraphics[width=209pt]{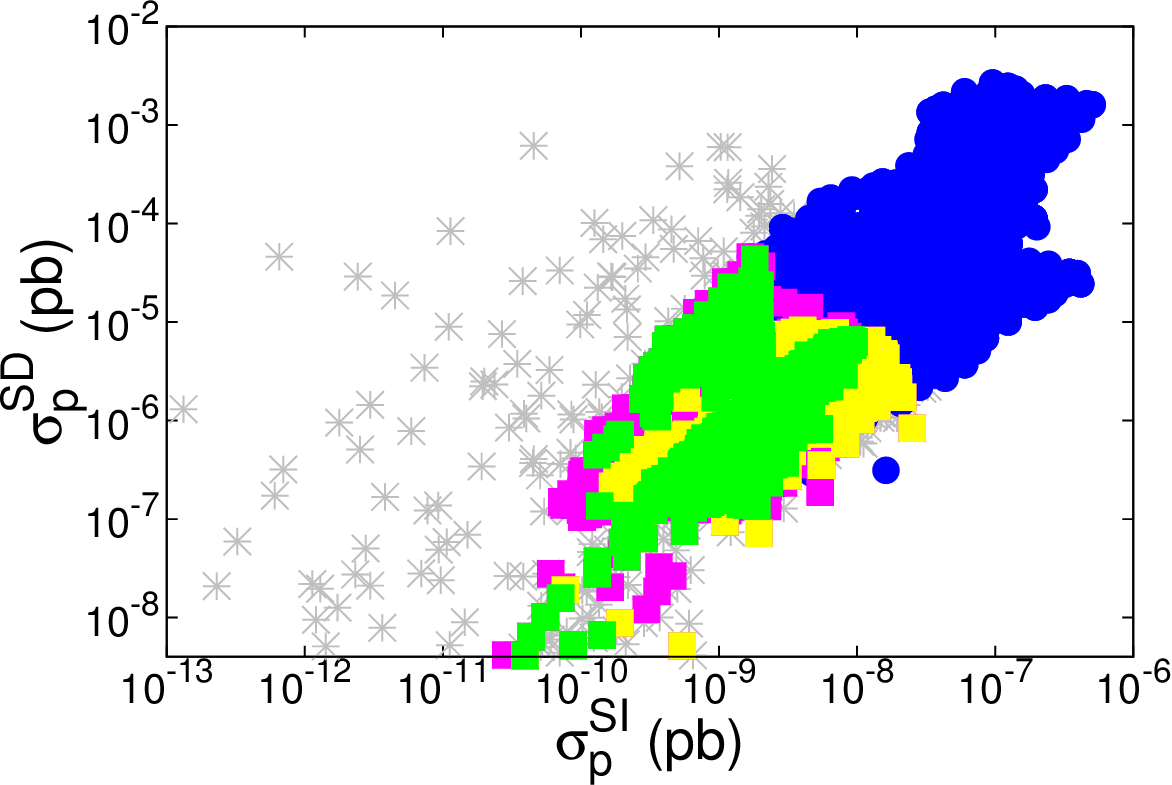}}
\caption[]{(a) the annihilation cross section $\langle \sigma_{\rm a} v \rangle (v\rightarrow0)$ versus the spin independent cross section $\sigma_P^{SI}$. (b) the spin-dependent cross section $\sigma_P^{SD}$ versus the spin independent cross section $\sigma_P^{SI}$. Legends are the same as in figure~\ref{fig:xenon}.}
\label{fig:svsi}
\end{figure}


The WIMP DM at the present epoch is non-relativistic and we can thus relate the current indirect search via  LSP annihilation to that at freeze out \cite{Barger:2008qd}. The partial wave properties of the LSP annihilation allow us to understand the various contributions. figure~\ref{fig:svsi}(a) shows the annihilation cross section $\langle \sigma_{\rm a} v \rangle (v\rightarrow0)$ versus the spin-independent cross section when scattering off a proton $\sigma_{\rm p}^{SI}$. The model points in green squares near half the $Z$ boson and near half the $126~\gev$ Higgs boson in figure~\ref{fig:cx}(a) correspond to the low branch of the green squares in figure~\ref{fig:svsi}(a), due to the p-wave suppression. On the other hand, the $s$-channel annihilation through $A$ in the mass window $200~\gev~\sim~500~\gev$ in figure~\ref{fig:cx}(a) is through $s$-wave, and thus has a relatively high cross section (indicated by the high branch of green squares). Although the LSP couplings to $H$ and $A$ both are mainly through their Higgsino components $N_{14}$, the $H$ exchange is via $p$-wave and thus yields a lower cross section as shown by the middle branch in figure~\ref{fig:svsi}(a). Finally, we note that the LSP-NLSP co-annihilation (yellow squares) could yield higher cross sections for both direct, and indirect searches, depending on their Wino and Higgsino components. figure~\ref{fig:svsi}(b) shows the spin-dependent cross section versus the spin-independent cross section, for our different models. Some of the models represented by blue points have a large enough cross section to be probed by IceCube/DeepCore \cite{icecube}. However, a large spin-dependent cross section implies a proportionally large spin-independent cross section. Thus, all models in blue are excluded by the XENON-100 experiment.  figure~\ref{fig:svsi} illustrates the connection between spin-dependent and spin-independent measurements, as well as the connection between direct searches and indirect searches.
Further improvement of the indirect search sensitivity will soon reach the relevant parameter region, and will probe the $A$ exchange contribution due to its $s$-wave dominance.
%

\subsection{Implication of LSP for Higgs physics}

\begin{figure}[t]
\begin{center}
\subfigure[]{
\includegraphics[width=209pt]{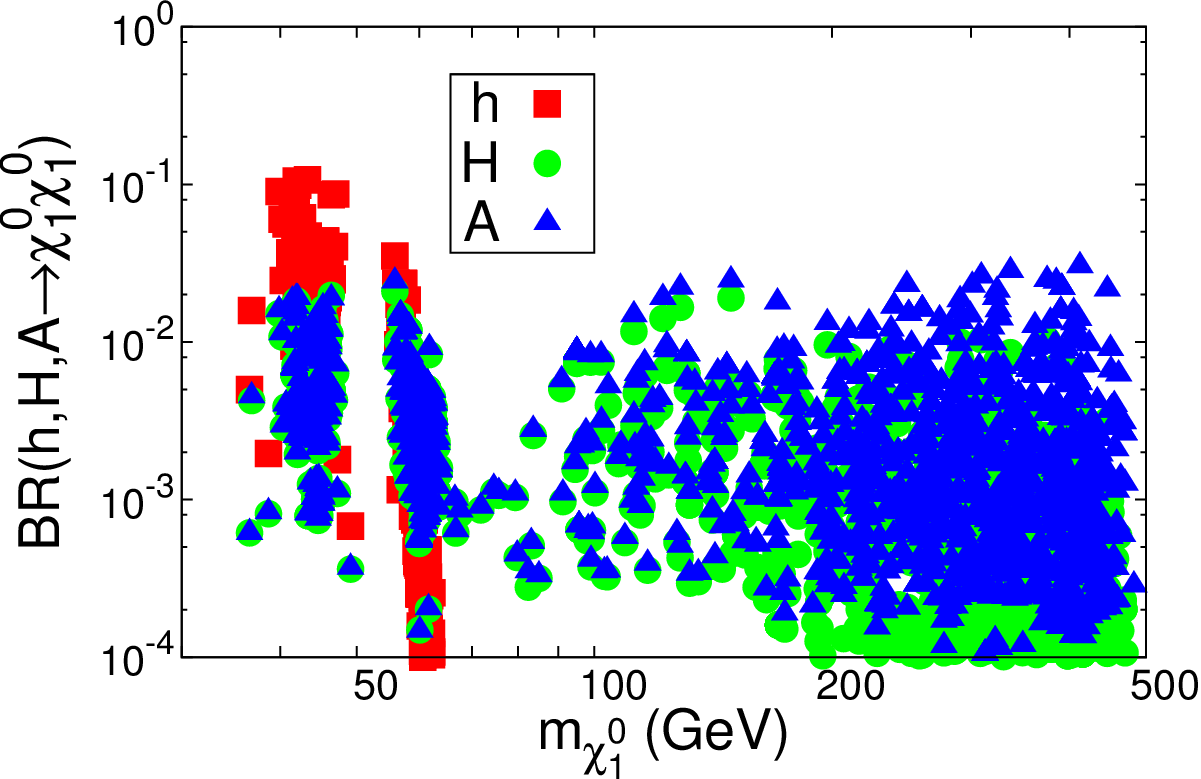}}
\subfigure[]{
\includegraphics[width=209pt]{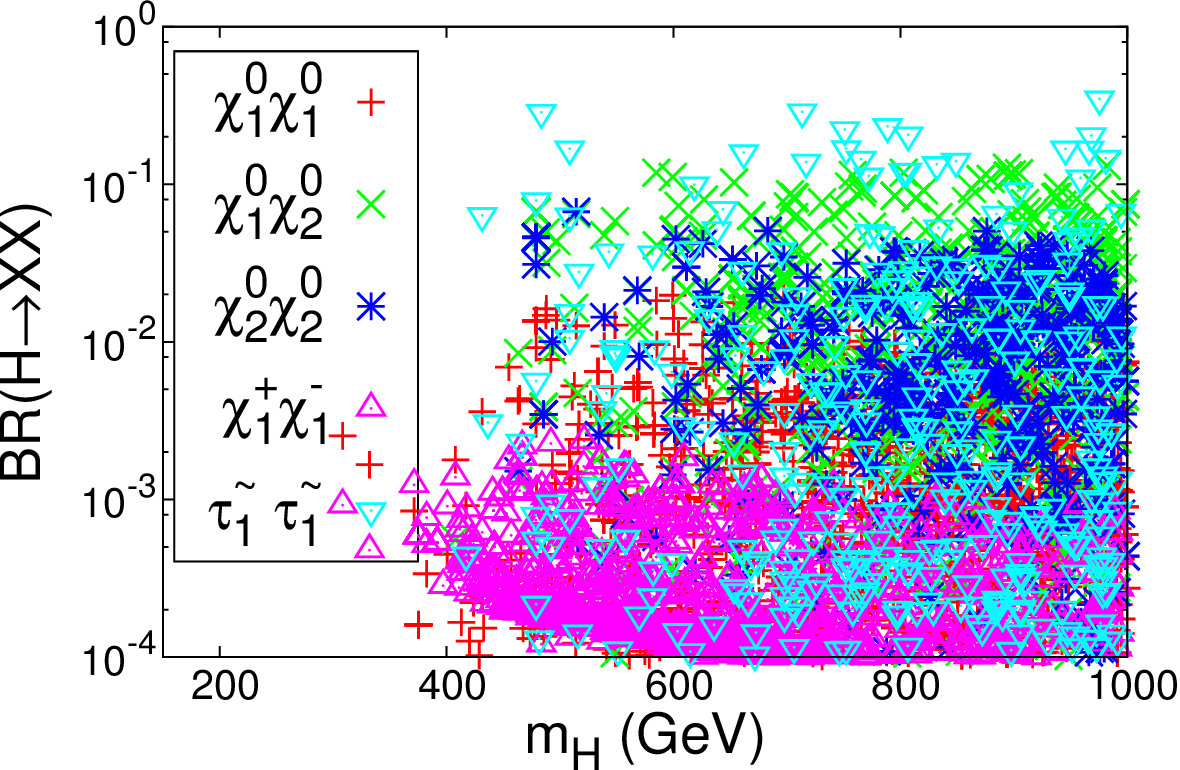}}
\subfigure[]{
\includegraphics[width=209pt]{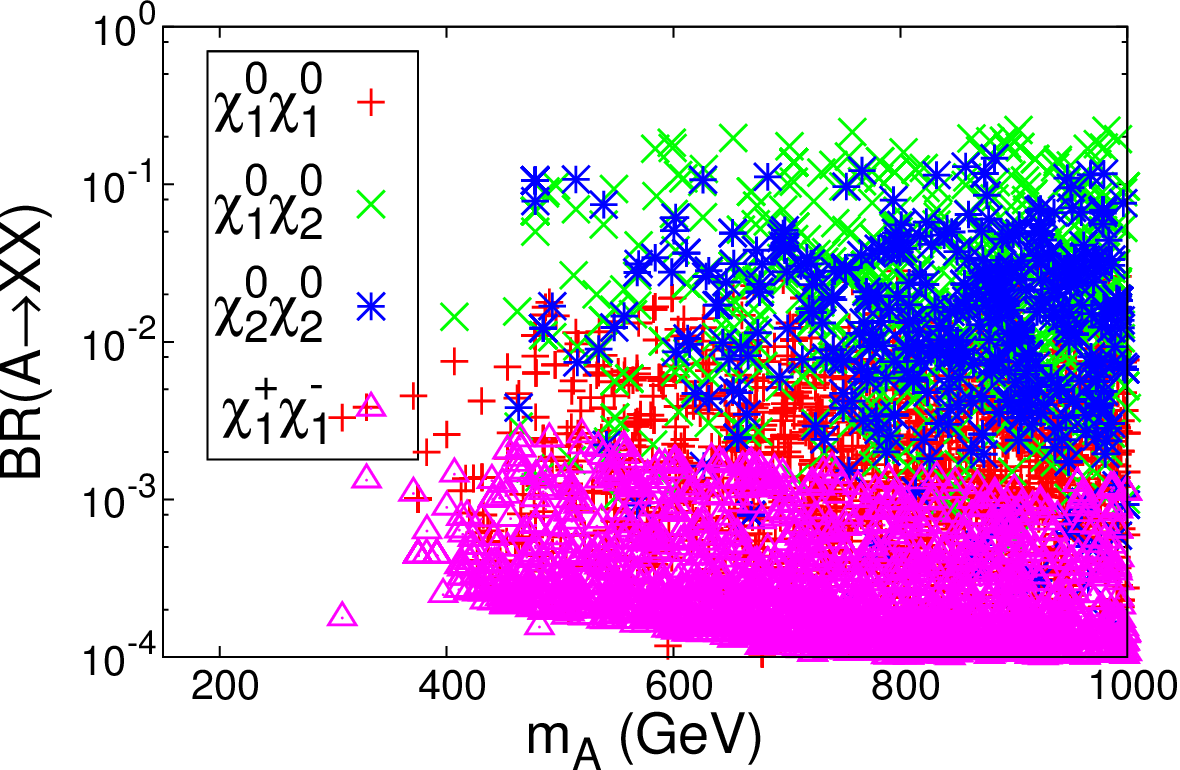}}
\subfigure[]{
\includegraphics[width=209pt]{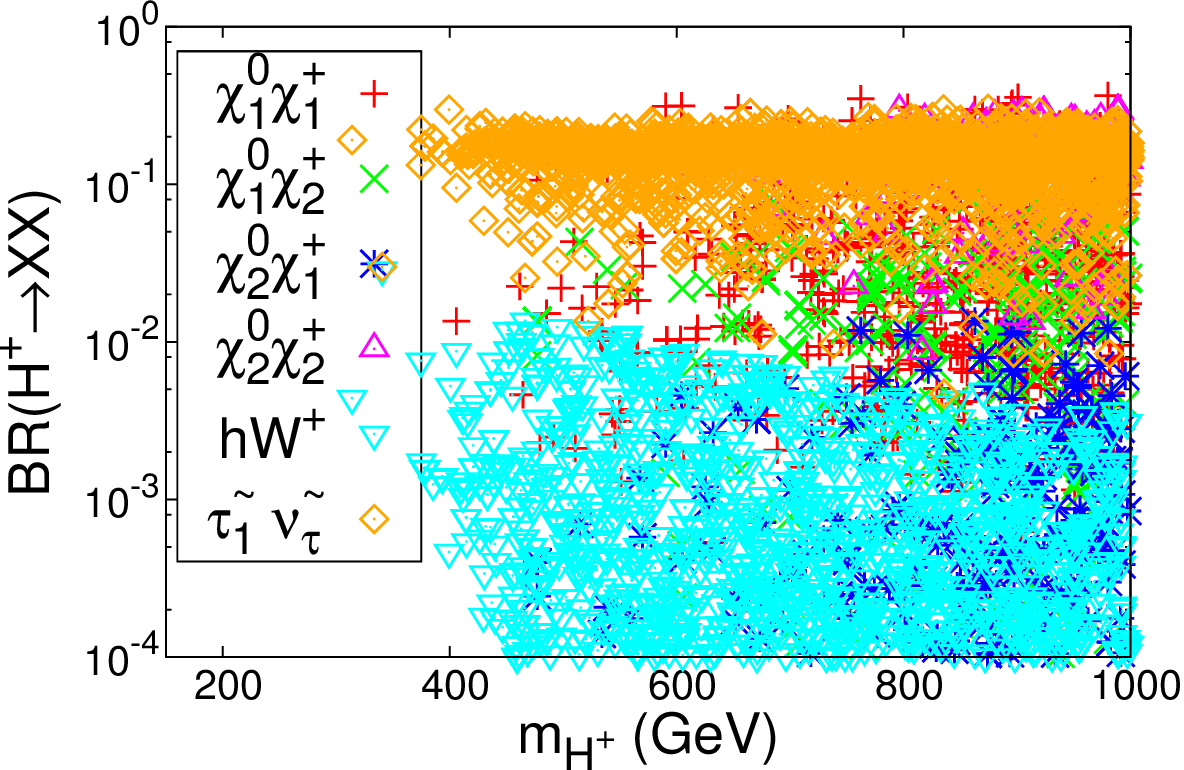}}
\end{center}
\caption[]{Branching fractions to neutralinos and charginos
(a) for $h, H, A$ decays to LSP pair versus the LSP mass,
(b) for $H$, (c) for $A$, and (d) for $H^{\pm}$ versus its mass respectively.
\label{fig:BRX}}
\end{figure}
As shown in figure~\ref{fig:relics}, a class of solutions exist with the LSP mass nearly half the mediator $Z,~h,~H, ~A$ mass that undergoes a resonant enhancement in annihilation, in the Higgs funnel region. One may expect to see the mediator's invisible decay mode to LSP pairs in collider experiments. Unfortunately, these channels are kinematically suppressed near threshold by the non-relativistic velocity factor. Near the $Z$ peak for example, the search for $Z\to \chi_{1}^{0} \chi_{1}^{0}$ would prove impossible since the branching fraction would be smaller than $10^{-5}$ due to this suppression. On the other hand, invisible decay channels could be sizable for heavier parent particles.
Shown in figure~\ref{fig:BRX}(a) are the branching fractions of $h,\ H,\ A$ to a pair of LSP $\chi^0_{1}\chi^0_{1}$  versus its mass, which would be the invisible mode in collider experiments. It is informative to note that the SM-like Higgs boson receives two distinctive contributions denoted by the red squares
\bea
{\rm BR}_{max}(h \to \chi^0_{1}\chi^0_{1})
& \sim &
\left\{
\begin{array}{ll}
1\%  & \quad  m_{\chi} \approx 60\ {\rm GeV}, \\
10\%  & \quad  m_{\chi} \approx 45\ {\rm GeV}.
\end{array}
\right.
\eea
The branching fraction near  $60~\gev$ is rather small although this is clearly identifiable as the $h$-funnel region. The branching fraction near  $45~\gev$ is about an order of magnitude larger because of the available kinematics, even though it is from the $Z$-funnel. This leads to the very interesting and challenging possibility of observing the Higgs invisible decay at the LHC \cite{Taoinvi,jetinvi,Gunioninvi}, (a sensitivity of about $20\%$ is considered feasible). The search sensitivity would be significantly improved at future $e^{+}e^{-}$ colliders, reaching about a few percent at the International Linear Collider (ILC), and even $0.3\%$ at the TLEP \cite{TLEP}.


\begin{figure}[t]
\begin{center}
\subfigure[]{
      \includegraphics[width=209pt]{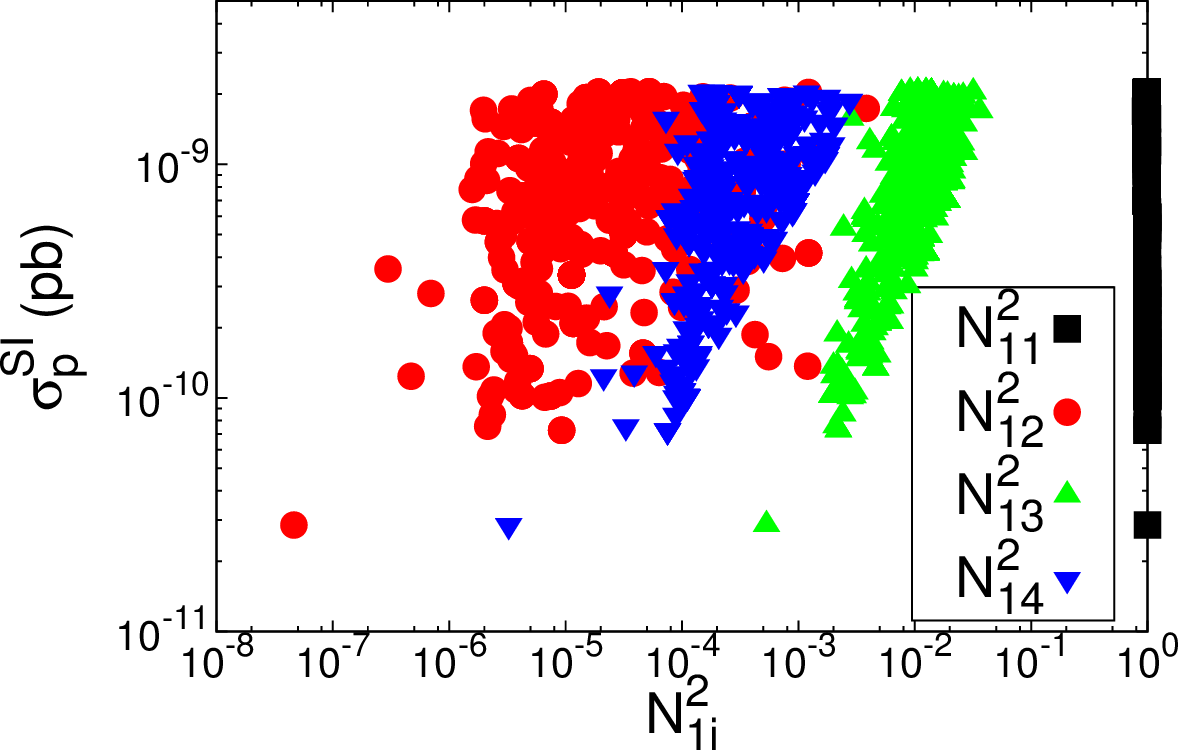}}
\subfigure[]{
      \includegraphics[width=209pt]{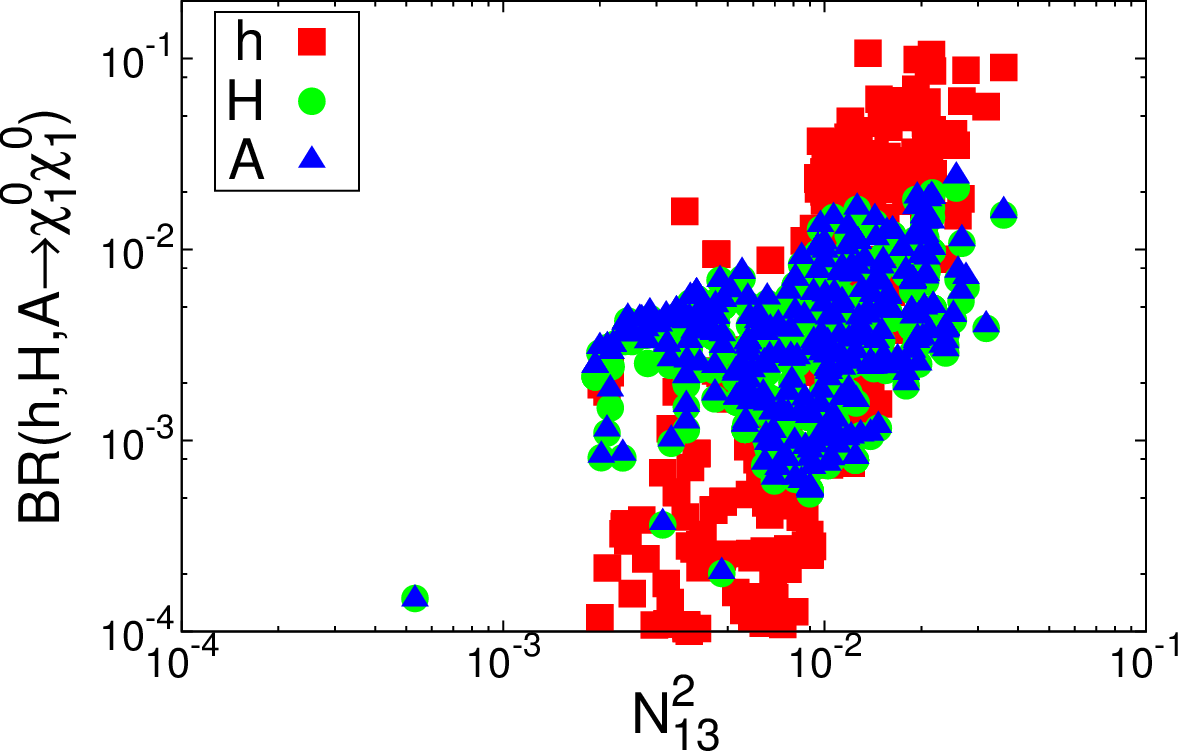}}
\subfigure[]{
      \includegraphics[width=209pt]{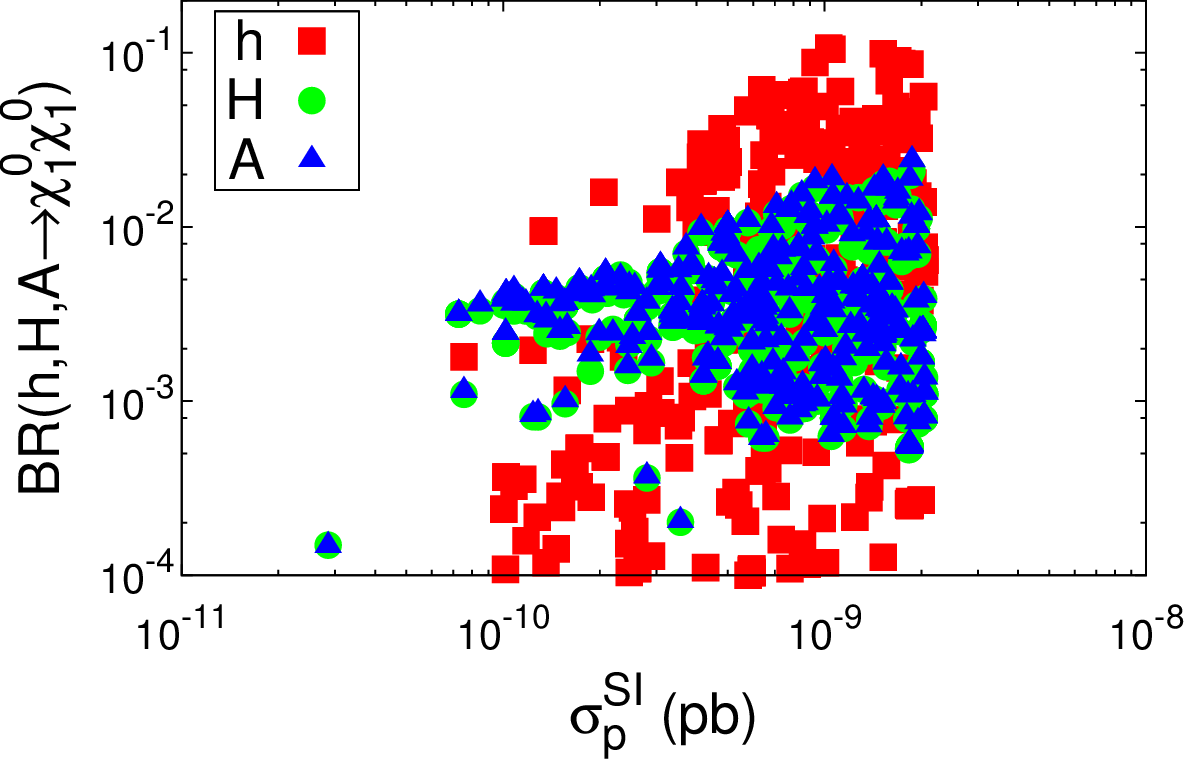}}
\subfigure[]{
      \includegraphics[width=209pt]{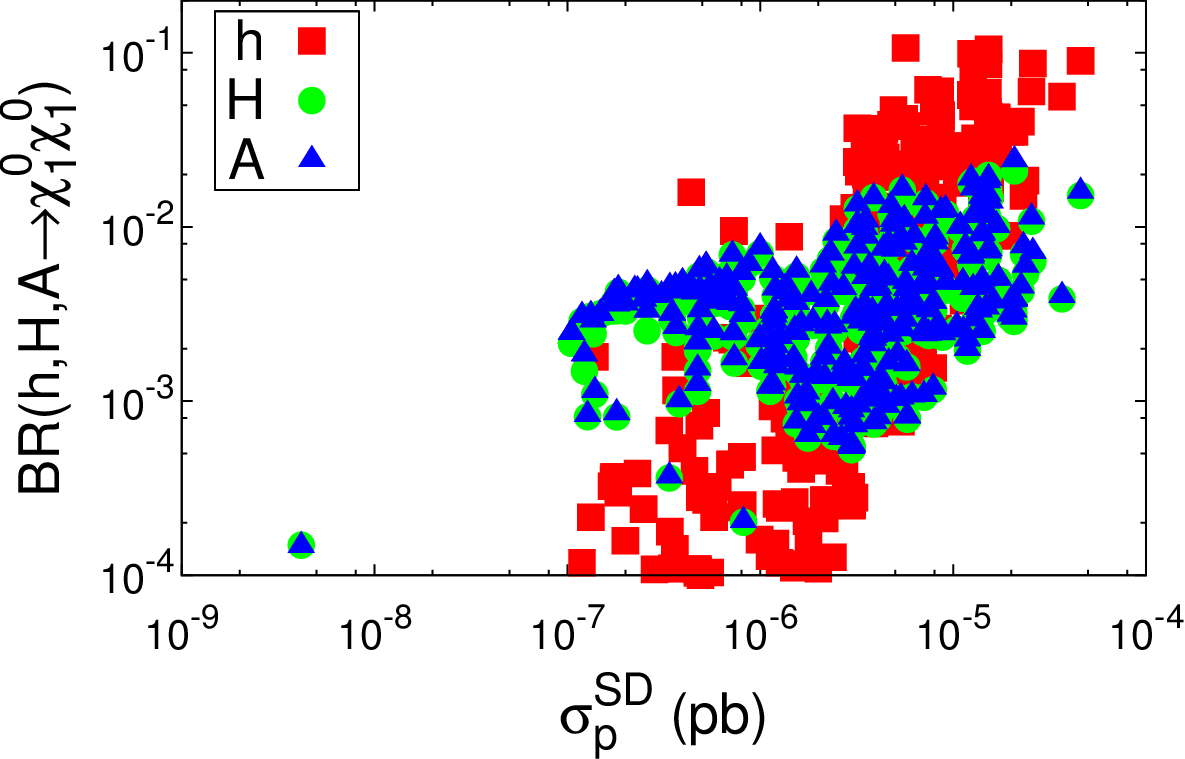}}
\end{center}
\caption[]{(a) Spin-independent cross section versus the gaugino and Higgsino fractions $N_{1i}^{2}$, and neutral Higgs decay branching fractions to DM pairs (b) versus the leading Higgsino fraction $N_{13}^{2}$, (c) versus spin-independent cross section, and (d) versus spin-dependent cross section.
}
\label{fig:zoom}
\end{figure}

\subsection{Consequences of co-annihilation}

For the co-annihilation scenarios, the other lower-lying SUSY particles are nearly degenerate with the LSP to ensure efficient annihilation. The common case is that the NLSP and NNLSP of the Winos ($\chi^{\pm}_{1},\ \chi^{0}_{2}$) or the Higgsinos  ($\chi^{\pm}_{1},\ \chi^{0}_{2,3}$) are nearly degenerate with the Bino-like LSP, with appreciable mixing among them. On the other hand, the XENON-100 search bound puts a constraint on the sizes of the mixing as seen from Fig~\ref{fig:fractions}(a) and (b). Nevertheless, the spin-independent cross sections are typically higher than those from the Higgs resonances, reaching $\sigma_p^{\rm SI} \sim 10^{-8}~{\rm pb}$ (yellow region). The indirect detection cross sections are in general between the $s$-wave dominance (higher green band) and $p$-wave dominance (lower green band).

As shown in Figs.~\ref{fig:BRX}(b)$-$(d), branching fractions for the other heavy Higgs bosons $H, A$ to a pair of light SUSY particles could reach up to about $10\% - 20\%$. These are the solutions for the correct relic density with co-annihilations. However,
due to the mass degeneracy, the final decay products would be rather soft and would be difficult to observe with the LHC. Consequently, these also yield the invisible decay channels.

The coannihilation scenarios predict a rich spectrum near the LSP mass, leading to many different phenomena that can be explored by sparticles pair productions \cite{Arganda:2012qp, Dutta:2013sta, SanjayHanSu}.

To conclude our discussion in this section, we bring a few crucial observables to comparison. First, in figure~\ref{fig:zoom}(a), we show the spin-independent cross section labelled by the gaugino components $N_{1i}^{2}$ of Bino (black), Higgsinos (green and blue), and Wino (red).
The lower right slopes of the $N_{13}^2$ and $N_{14}^2$ regions in this plot indicate the variable contributions from $H$-exchange and $h$-exchange, respectively, as discussed earlier in eq.~(\ref{eq:hh}).
 %
We then show the Higgs decay branching fractions versus the leading gaugino component $N_{13}^{2}$ in figure~\ref{fig:zoom}(b). We see that the higher branching fractions naturally correspond to a higher value of the mixing parameter. In Figs.~\ref{fig:zoom}(c) and (d), we reiterate the correlations among the observables by showing the neutral Higgs decay branching fractions versus spin-independent cross section and spin-dependent cross section, respectively. It is a generic feature that higher Higgs decay branching fractions correspond to higher cross sections. It is interesting to see that the spin-dependent cross section shows slightly more correlation with the Higgs BR's. We see the similarity between Figs.~(b) and (d). This comes from the fact the $Z$-exchange in spin-dependent cross section is governed by $N_{13}^2$ while $N_{14}^2$ is rather small.
It is important to emphasize that in anticipation of the improvement of the direct search in the near future, the LUX and XENON-1T experiments would be able to cover the full parameter space, pushing down to very small Higgs branching fractions, as shown in  figure~\ref{fig:zoom}(c).


\section{Summary and conclusions}
\label{Conclude}

\begin{table}[t]
\begin{center}
\begin{tabular}{|c|c|c|c|c|c|c|c|}
  \hline
  Type  & DM mass & Annihilation & Partial & $\langle \sigma v \rangle (v\to 0)$ & Collider \\
   labels  & $m_{\chi_1^0}$ &  channels & waves &  & searches \\ \hline
  I-A & $\sim m_Z/2$ & $\ccto Z  $ & p & low & $Z, h, H, A\to \chi_1^0\chi_1^0$\\ \hline
  I-B &  $\sim m_h/2$ & $\ccto h  $ & p & low & $h, H, A\to \chi_1^0\chi_1^0$\\ \hline
  I-C &  $\sim m_A/2$ & $\ccto A  $ & s & high & $H, A\to \chi_1^0\chi_1^0$\\ \hline
  & $m_{\chi_1^0}\sim m_{\chi_1^\pm}$ & $\chi_1^0 \chi_2^0, \chi_1^0 \chi_1^\pm$ & & & $H,A\to \chi_1^0\chi_2^0$\\
  II-A  & $\sim m_{\chi_2^0}$ & $\chi_2^0 \chi_2^0, \chi_1^+ \chi_1^-$ & s+p & medium & $H,A\to \chi_2^0\chi_2^0$\\
  &  & $\to SM$ & & & $H^\pm\to \chi_1^0\chi_1^\pm$\\ \hline
  & $m_{\chi_1^0}\sim m_{\tilde \tau_1}$ & $\tilde \tau_1^+ \tilde \tau_1^-, \tilde \nu_\tau \tilde \nu_\tau,$ & & & $H,A\to\tilde \tau_1^+ \tilde \tau_1^-$\\
  II-B & $\sim m_{\tilde \nu_\tau}$  & $\chi_0^1 \tilde \tau_1^\pm \to SM$ & s+p & medium & $H^\pm \to \tilde \tau_1^\pm \tilde \nu_\tau$\\ \hline
  \hline
\end{tabular}
\caption[]{Connection between the SUSY DM properties and the Higgs bosons.}
\label{tab:summary}
\end{center}
\end{table}

Within the framework of the minimal supersymmetric extension of the standard model (MSSM), we investigated the possibility of the lightest supersymmetric particle being all the dark matter in light of the recent discovery of a SM-like Higgs boson, and the search for other Higgs bosons and SUSY particles at the LHC. We scanned through a wide range of the MSSM parameter space, and searched for model points wherein LSP has the correct properties to be the (WIMP) thermal DM.
We studied the freeze-out of WIMPs and computed the cross section required to give the correct DM relic abundance in terms of velocity-independent and velocity-dependent components, as shown in figure~\ref{fig1}.
We applied the constraints on the MSSM Higgs sector from the LEP, Tevatron and LHC observations. We also imposed flavor constraints from the recent experimental results at the LHCb and BELLE, and found stringent bounds on the parameter space. The low LSP mass region may be closed, yielding a rough bound $m_\chi > 30~\gev$, unless for a compressed SUSY spectrum such as $m_{\tilde b}-m_{\chi}<5~\gev$.

The XENON-100 experiment significantly constrains the viable parameter region via the spin-independent elastic WIMP-proton scattering cross section, as shown in Figs.~\ref{fig:para}-\ref{fig:xenon}. Although not as sensitive, the indirect search experiments such as Fermi/LAT and IceCube have obtained impressive results to cut into the SUSY parameter region, as seen in figure~\ref{fig:cx}.
We are able to identify the Higgs contributions and thus to make predictions for future searches at the LHC and ILC. There are also clear contributions from the co-annihilation channels. Table~\ref{tab:summary} summarizes these distinctive MSSM model points, and the relation with the Higgs bosons. We reiterate the key points of our findings. For the resonance scenarios as in I-A, I-B and I-C,
\begin{itemize}
\item $Z,\ h,\ H$ and $A$ are the most important mediators at resonance to yield the correct relic abundance and give predictive narrow mass windows as shown in figure~\ref{fig:xenon}, which we refer to as the $Z$, $h$ and $H/A$-funnel regions.
The spin-independent scattering in the $Z,h$ funnel is dominated by the $t$-channel $H$ exchange when $N_{13}^2 \gg N_{14}^2$, which is mostly the case seen in figure~\ref{fig:fractions}(b).
\item With our parameter scanning, the necessarily non-zero Wino, Higgsino components of the LSP (as seen in figure~\ref{fig:fractions}) imply a lower bound for the WIMP scattering cross section mediated by $h$ and $H$, as in eq.~(\ref{eq:bound}). In particular, the spin-independent cross sections may be fully covered by the next generation of direct search experiments for DM mass around $30-800~\gev$ such as LUX and XENON-1T, as seen in figure~\ref{fig:xenon} and figure~\ref{fig:zoom}(c).
An exception is the fine-tuned cancellation, the ``blind spots'' scenario, way above the $Z,h$ funnels, as shown by the grey crosses in figure~\ref{fig:xenon}.
\item $Z,\ h,\ H$ and $A$ mediators determine the partial wave decomposition as listed in Table \ref{tab:summary} and predict a definite range of indirect search cross sections. It is especially sensitive to the $A$-exchange contribution, as seen in figure~\ref{fig:svsi}(a).
\item The invisible decays of $h,\ H$ and $A$ are expected, as plotted in figure~\ref{fig:BRX}(a). Future studies at the LHC, and in particular, at the ILC may reveal the true nature of the DM particle, as seen in Figs.~\ref{fig:zoom}(b-d).
\end{itemize}

\noindent
For the co-annihilation scenarios as in II-A and II-B,
\begin{itemize}
\item
Although the ``well-tempered'' scenario with large Higgsino and Wino fractions is disfavored by the \xen data, the co-annihilation may still be a valid solution to obtain the correct relic density.
There may be several light SUSY particles such as neutralinos, charginos, or stau, leading to many rich phenomena that can be searched for at the LHC, and may be fully covered by the ILC.
%
\item For highly degenerate NLSP, NNLSP, the decays of $H,\ A$ and $H^\pm$ as shown in Figs.~\ref{fig:BRX}(b)-(d) could lead to large invisible modes, making the collider search for DM very interesting.
\end{itemize}

 We conclude that understanding the nature of DM requires us to consider results from a number of different experiments. Future collider searches and the next generation of direct detection experiments will likely cover  the conventional parameter range of the MSSM if the LSP constitutes all of the DM.  The recent exciting discovery of the SM-like Higgs boson, and searches for beyond the SM physics at the energy frontier will serve as a new ``lamp post'' and guide in DM searches complementary to what may be obtained from direct detection and astro-particle observations at the cosmic frontier.

\acknowledgments{We would like to thank M.~Drees, B.~Dutta, G.~Gelmini, J.~Fan, L. Roszkowski, J.~Ruderman, X.~Tata and L.-T. Wang for helpful discussions. In particular, we thank Shufang Su for many helpful comments and discussions, which resulted in adding figure~\ref{fig:zoom} to visualize the correlations among the observables. T.H.~and Z.L.~are supported in part by the U.S.~Department of Energy under Grant No. DE-FG02-95ER40896, in part by the PITT PACC. Z.L.~is also supported in part by the LHC Theory Initiative from the U.S. National Science Foundation under Grant No. NSF-PHY-0969510. A.N.~is supported in part by a McWilliams postdoctoral fellowship awarded by the Bruce and Astrid McWilliams Center for Cosmology, and in part by NSF grant AST-1211777. T.H.~would also like to thank the Aspen Center for Physics for the hospitality during which part of the work was carried out.  The Aspen Center for Physics is supported by the NSF under Grant No.1066293.
}

\appendix

\centerline{\bf Appendix}

\section{Relic density calculation}
\label{App:freeze_out}

When the Hubble expansion $H = \dot a / a$ became much larger than the interaction rate $\Gamma = n_\chi \sig$, the WIMPs ($\chi^{0}$), once in thermal equilibrium with the rest of the Universe, decoupled from equilibrium.
The number density of WIMPs at a time $t$, $n_\chi (t)$, is obtained by solving the Boltzmann equation
\beq
\frac{1}{a^3} \, \frac{d \left( a^3 n_\chi \right )}{dt} = -\sig \left [ n^2_\chi - n^2_{\rm eq} \right ],
\label{bolt1}
\eeq
where $\sig$ is the WIMP annihilation rate averaged over velocities, and $n_{\rm eq}$ is the equilibrium number density of WIMPs:
\beq
n_{\rm eq} = \frac{g}{2 \pi^2} \;  \int_{m_\chi}^\infty \; dE \; \frac{E \sqrt{E^2 - m^2_\chi}}{1 + e^{E/T}}.
\eeq
$g$ measures the number of relativistic degrees of freedom, and $T$ is the temperature.

Define the dimensionless variables $Y = n_\chi / s$ and $x = m_\chi / T$, where $s$ is the entropy density given by
\beq
s = \frac{2\pi^2}{45} \, g_{\rm s} T^3.
\label{entropy2}
\eeq
Here and henceforth, we adopt the natural units $k_{\rm b} = \hbar = c = 1$. $g_{\rm s}$ is different from $g$ only at late times, after neutrinos have decoupled from equilibrium, and $e^\pm$ annihilation leads to the photons being heated relative to the neutrinos. However, $dY/dx$ is very small at late times, and with  good accuracy, we may set $g = g_{\rm s}$ when computing the relic abundance. The entropy per comoving volume is conserved, and hence
\beq
\frac{d (a^3 s)} {dt} = \frac{d \left( g a^3 T^3 \right ) }{dt} = 0
\label{entropy1}
\eeq
We may then rewrite eq.~(\ref{bolt1}) in terms of $Y$ as:
\beq
\frac{dY}{dt} = -\sig s \left [ Y^2 - Y^2_{\rm eq} \right ],
\label{eq:Y1}
\eeq
where $Y_{\rm eq} = n_{\rm eq} / s$. From eq.~(\ref{entropy1}), we see that $g a^3 T^3$ = constant, and therefore,
\beq
-\frac{\dot T}{T} = H + \frac{\dot g}{3g}.
\eeq
The Hubble parameter $H(T)$ is given by the expression
\beq
H = \left [ \frac{8 \pi G }{3 } \; \rho_\gamma \right ]^{1/2} = \left [ \frac{8 \pi^3 G }{90 } \right ]^{1/2} g^{1/2} \; T^2
\label{hubble}
\eeq
Differentiating $x=m_\chi/T$ with respect to time, we find
\bea
\dot x = \left(-\dot T / T \right ) x &=& Hx \, \left [1 + \frac{\dot g_s}{3Hg_{\rm s}} \right ] \n
&\approx& Hx \left [ 1 - \frac{1}{3} \frac{d(\ln g_{\rm s})}{d(\ln T)} \right ],
\label{xt}
\eea
where we simplified the second term on line 1 by substituting $\dot x \approx Hx$ (provided $\dot g_{\rm s} \ll 3 H g_{\rm s}$), and therefore $\dot g \approx -HT (dg / dT)$ \cite{beacom}. Note that $g$ changes significantly at the epoch of quark-hadron transition. We may now rewrite eq.~(\ref{eq:Y1}) in terms of $x$:
\bea
\frac{dY}{dx} &=& \frac{ -\sig s(x) }{H(x) x \left [ 1 - \frac{1}{3} \frac{d(\ln g)}{d(\ln T)} \right ]} \n
&=& - \sqrt{ \frac{\pi } {45 G }} \; \sig m_\chi \; \frac{ g^{1/2}  }{ \left [ 1 - \frac{1}{3} \frac{d(\ln g)}{d(\ln T)} \right ] }   \; \frac{ Y^2 - Y^2_{\rm eq} }{x^2}.
\label{bolt_equation}
\eea

We solve eq.~(\ref{bolt_equation}) numerically to obtain the present day value $Y_0$, once the form of $g(T)$ and $\sig$ are known. The dark matter relic density is then computed as:
\beq
\rel = \frac{m_\chi Y_0 s_0}{ \left( \rho_{\rm crit} / h^2 \right ) },
\label{omchih2}
\eeq
where $s_0 \approx 2893$ cm$^{-3}$ is the present day entropy density and $\rho_{\rm crit} \approx 1.054 \times 10^{-5}  \, h^2$ GeV/cm$^{3}$ is the critical density.

\begin{figure}[t]
\hspace{0.25in} (a) \hspace{2.7in}  (b)
\begin{center}
\scalebox{0.50}{\includegraphics{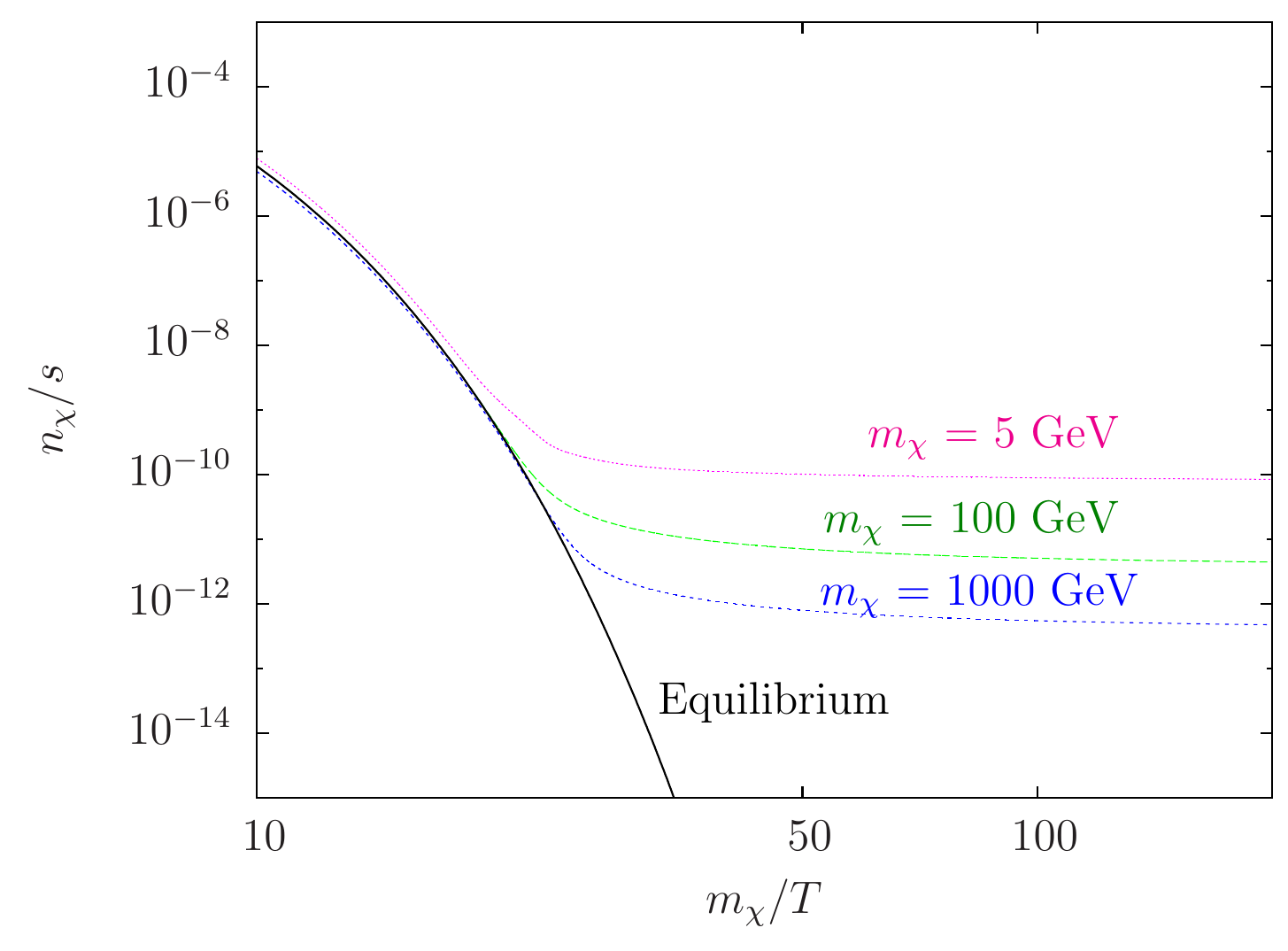}}
\scalebox{0.38}{\includegraphics{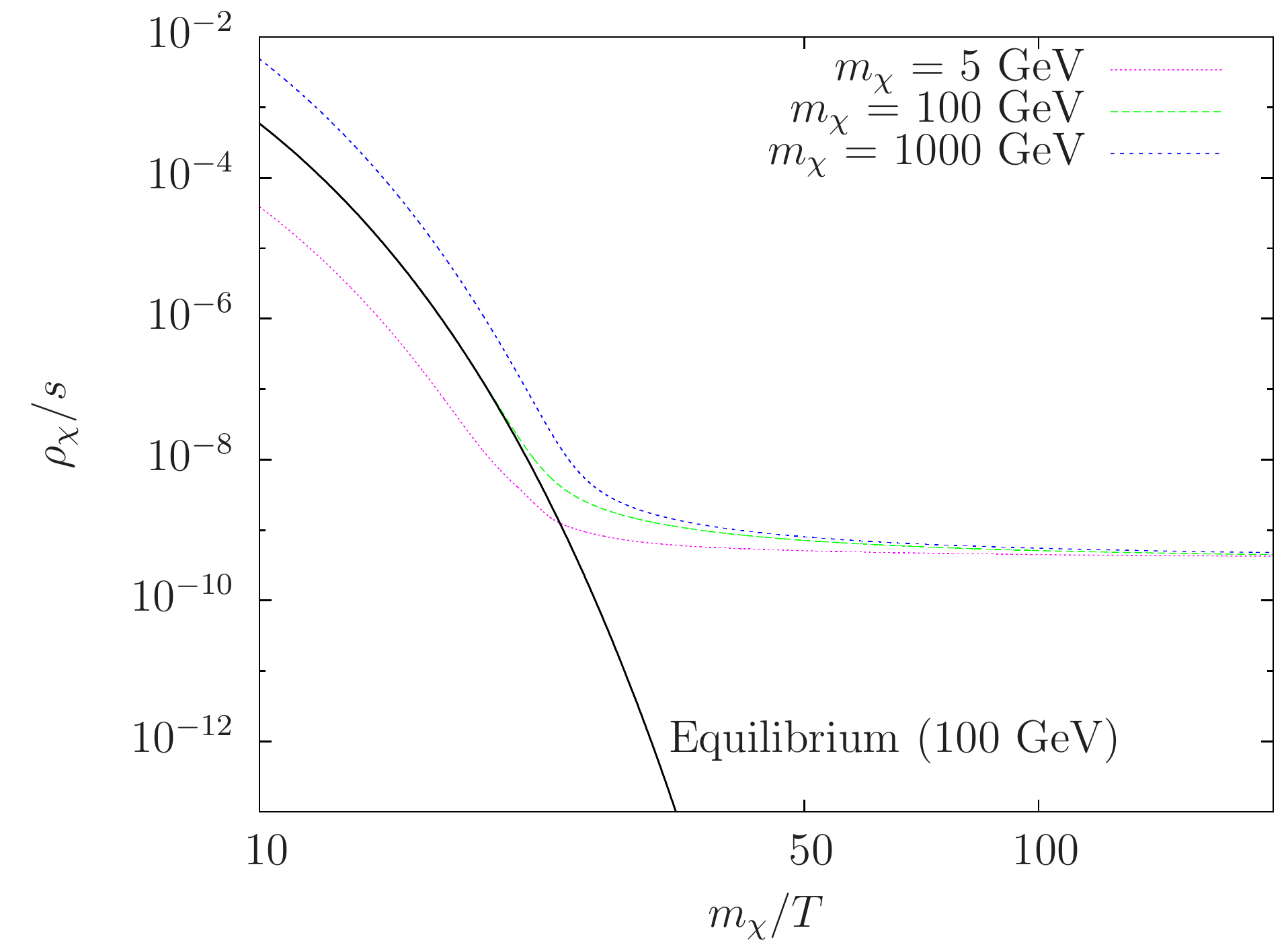}}
\end{center}
\caption[]{
Evolution with temperature and yielding the correct WIMP relic density $\rel$  = 0.11 with illustrative values of the WIMP mass $m_\chi = 5,$ 100, and 1000 GeV,
(a) WIMP number density, and (b) WIMP mass density. The equilibrium lines are for $m_\chi = 100$ GeV.
\label{fig:relics} }
\end{figure}
After performing a numerical integration of the Boltzmann equation as formulated in eq.~(\ref{bolt_equation}), we show the WIMP number density in figure~\ref{fig:relics}(a) and the WIMP relic (mass) density in figure~\ref{fig:relics}(b), for various WIMP mass values. The dark straight-falling line gives the densities if the particle keeps in thermal equilibrium with the environment for $m_{\chi}=100$ GeV.
It is known that the freeze-out temperature for a relatively light WIMP particle is
\beq
x_{\rm f} = m_{\chi}/T \approx 20.
\eeq
The horizontal curves in Figs.~\ref{fig:relics}(a) and (b) present the WIMP number density and mass density after freeze-out for $m_{\chi}=5-1000$ GeV, leading to the correct relic density.

\bibliographystyle{JHEP}
\bibliography{references}

\end{document}